\documentclass[12pt,notitlepage]{revtex4}

\usepackage{graphicx}
\usepackage{epstopdf}

\font\msytw=msbm9 scaled\magstep1
\font\msytww=msbm7 scaled\magstep1

\font\cs=cmcsc10

\let\a=\alpha \let\b=\beta    \let\d=\delta \let\e=\varepsilon
\let\z=\zeta  \let\h=\eta   \let\th=\theta \let\k=\kappa 
             \let\p=\pi    
\let\s=\sigma \let\t=\tau   \let\f=\varphi 
   
 \let\D=\Delta

\def\EE{{\cal E}} 
\def\FF{{\cal F}} \def\HHH{{\cal H}}
\def\NN{{\cal N}} \def\III{{\cal I}}

\def\RRR{\hbox{\msytw R}}

 \def\ZZZ{\hbox{\msytw Z}}
\def\zzzz{\hbox{\msytww Z}} 

\def\\{\hfill\break} 
\let\==\equiv
\let\io=\infty 
\let\0=\noindent

\let\dpr=\partial
\def\sign{{\rm sign}}
\def\const{{\rm const}}
\def\tende#1{\,\vtop{\ialign{##\crcr\rightarrowfill\crcr\noalign{\kern-1pt
    \nointerlineskip} \hskip3.pt${\scriptstyle #1}$\hskip3.pt\crcr}}\,}
\def\otto{\,{\kern-1.truept\leftarrow\kern-5.truept\to\kern-1.truept}\,}

\def\to{\rightarrow}

\def\qed{\hfill\raise1pt\hbox{\vrule height5pt width5pt depth0pt}}

\def\be{\begin{equation}}
\def\ee{\end{equation}}
\def\bea{\begin{eqnarray}}
\def\eea{\end{eqnarray}}
\def\nn{\nonumber}

\begin{document}

\title{Striped periodic minimizers of a two-dimensional model for martensitic phase 
transitions}

\author{Alessandro Giuliani}\affiliation
{Dipartimento di Matematica di Roma Tre, Largo S. Leonardo Murialdo 1,
00146 Roma, Italy.}
\author{Stefan M\"uller}
\affiliation{Hausdorff Center for Mathematics \& Institute for Applied 
Mathematics, Universit\"at Bonn,
Endenicher Allee 60, 53115 Bonn, Germany.}

\begin{abstract} In this paper we consider a simplified two-dimensional scalar model for the
formation of mesoscopic domain patterns in martensitic shape-memory alloys at
the interface between a region occupied by the parent (austenite) phase and a 
region occupied by the product (martensite) phase, which can occur in two 
variants (twins). The model, first proposed by Kohn and M\"uller \cite{KM}, is 
defined by the following functional:
$$\EE(u)=\b||u(0,\cdot)||^2_{H^{1/2}([0,h])}+
\int_{0}^{L} dx \int_0^h dy\, \big( |u_x|^2 + \e|u_{yy}| \big)$$
where $u:[0,L]\times[0,h]\to\RRR$ is periodic in $y$ and $u_y=\pm1$ almost 
everywhere. Conti \cite{C06} proved that if $\b\gtrsim\e L/h^2$ then the 
minimal specific energy scales like $\sim \min\{(\e\b/L)^{1/2},
(\e/L)^{2/3}\}$, as $(\e/L)\to0$. In the regime $(\e\b/L)^{1/2}\ll
(\e/L)^{2/3}$, we improve Conti's results, by computing exactly the minimal 
energy and by proving that minimizers are periodic one-dimensional 
sawtooth functions. 
\end{abstract}

\maketitle

\renewcommand{\thesection}{\arabic{section}}

\section{Introduction and main results}\label{sec1}
\setcounter{equation}{0}
\renewcommand{\theequation}{\ref{sec1}.\arabic{equation}}

The formation of mesoscopic scale patterns in equilibrium systems is often due
to a competition between interactions favoring different microscopic 
structures; e.g., a competition between a short range attractive 
interaction favoring a homogeneous ordered state and a long range repulsive 
interaction, which opposes such ordering on the scale of the whole sample. 
Mathematically, this phenomenon can be modeled by (non-convex) 
free-energy functionals, whose minimizers are supposed to describe the low 
energy states of the system. The details of the free-energy 
functional to be considered depend on the specific system one wants to 
describe: applications range from micromagnetics \cite{DKMO,CKO99,GD82,MWRD95} 
to diblock copolymers \cite{ACO09, BF99, Ch01, RW05}, 
elasto-plasticity \cite{CO05, BCDM02}, 
superconducting films \cite{CCKO08, EK93, SK04} 
and martensitic phase transitions \cite{CO00, C06, KM, KM94}, 
just to mention a few. 

In all these cases, combinations of variational estimates and numerical 
simulations typically allow one to construct an approximate (and quite 
realistic) low temperature phase diagram, which often displays a 
wide range of ordering effects including formation of striped states 
\cite{KP93, Mu02, SS99},
droplet patterns \cite{GD82, Mu09}, triangular lattices 
\cite{A57, W34}, etc. However, a satisfactory theory of pattern formation 
in more than one dimension is still missing and the number of physical models 
for which periodicity can be rigorously proven is very small \cite{AM01,BL,CO, 
GLL1,GLL2,GLL3,GLL4, M, Suto, Th}. 
In this paper we prove periodicity of the minimizers of an anisotropic 2D 
free-energy functional, motivated by the theory of martensitic phase 
transitions. Our methods are based on a combination of reflection positivity 
estimates, in the spirit of \cite{GLL1,GLL2,GLL3,GLL4}, and of Poincar\'e-type 
estimates. We hope that these techniques will lead to more general examples of 
spontaneous pattern formation in anisotropic systems with competing 
interactions.

\section{Definition of the model and main results}\label{sec1000}
\setcounter{equation}{0}
\renewcommand{\theequation}{\ref{sec1000}.\arabic{equation}}

We consider a simplified two-dimensional (2D) scalar model for the formation
of mesoscopic domain patterns in martensitic shape-memory alloys at
the interface between a region occupied by the parent (austenite) phase and a 
region occupied by the product (martensite) phase, which can occur in two 
variants (twins). The model, first proposed by Kohn and M\"uller \cite{KM}, is 
defined by the following functional:
\be\EE(u)=
\b||u(0,\cdot)||^2_{H^{1/2}([0,h])}+
\int_{0}^{L} dx \int_0^h dy\, \big( |u_x|^2 + \frac{\e}2|u_{yy}| \big)
\label{1.1}\ee
where $u:[0,L]\times[0,h]\to\RRR$ is periodic in $y$ and $u_y=\pm1$ almost 
everywhere; therefore, the admissible functions are such that, for almost
every $x$, the graph of $y\to u(x,y)$ looks like a (possibly irregular) 
sawtooth pattern. Here $\b$ and $\e$ are nonnegative parameters. 
The square of the $H^{1/2}$-norm in the r.h.s. of (\ref{1.1}) is defined as:
\be ||u(0,\cdot)||^2_{H^{1/2}([0,h])}=4\p^2
\sum_{k\in \zzzz}|k||\hat u_0(k)|^2\;,\label{1.2}\ee
where $\hat u_0(k)=h^{-1}\int_0^h\,dy\, u(0,y)
\,e^{-2\p iky/h}$. The equivalent
$x$-space representation of this norm is the following:
\be ||u(0,\cdot)||^2_{H^{1/2}([0,h])}=\int_0^h dy \int_{-\io}^{+\io} dy'\,
\frac{|u(0,y)-\tilde u(0,y')|^2}{|y-y'|^2}\;,\label{1.3}\ee
where $\tilde u(0,y):\RRR\to\RRR$ is the periodic extension of $u(0,y)$ 
over the whole real axis. 

The problem consists in determining the minimizers of (\ref{1.1}) 
for small values of $\e$; existence of the minimizer was proved in \cite{KM94}.
As discussed in \cite{KM}, the significance of the various terms in (\ref{1.1})
is the following.
The rectangle $[0,L]\times[0,h]$ is the ``martensite'' region. The regions
where $u_y=-1$ and $u_y=1$ correspond to two distinct variants, 
which are separated by sharp interfaces. The term 
$\int|u_x|^2$ is the ``strain energy''; note that it vanishes 
only if the interfaces between the two twin variants are 
precisely horizontal, i.e., if the two variants form a striped (lamellar)
pattern. The term $(\e/2)\int |u_{yy}|$ is the surface energy;
since $u_y$ jumps from $-1$ to $+1$, $|u_{yy}|$ is like a delta function 
concentrated on the interfaces between the twins. It can be expressed 
more conventionally as 
\be \frac\e2\int_0^L dx\int_0^h dy |u_{yy}|=\e\int_0^L dx\, N(x)\;,\label{1.4}
\ee
where $N(x_0)=(1/2)\int_0^h dy |u_{yy}|$ is the number of twin boundaries that 
cross the line $x=x_0$. More precisely, $N(x)$ is defined as 
\be N(x)=\frac12 \sup\Big\{\int_0^h dy\, u_y(x,y) \f'(y)\;:\;
\f\in C^\io_{per}\ {\rm and}\ |\f|\le1\Big\}\;,\label{1.4a}\ee
where $C^\io_{per}$ is the set of periodic $C^\io$ functions on $[0,h]$. 
Note that if $\int\int u_x^2<\io$ then $x\mapsto u(x,\cdot)$ is a continuous map
from $[0,L]$ to $L^2([0,h])$ and, therefore, $N(x)$ is lower semicontinuous,
being a supremum of continuous functions. Note also that 
$\EE(u)<\io$ and the fact that $u_y\in L^\io([0,L]\times[0,h])$ imply that 
$u$ has a (1/3)-H\"older continuous representative \cite{R80}; therefore, 
in the following, with no loss of generality, we shall assume $u$ to be 
continuous in $[0,L]\times[0,h]$.

The boundary $x=0$ represents the interface between the martensite and the 
austenite and the term proportional to the square of the $H^{1/2}$-norm 
of $u(0,y)$ is the ``elastic energy in the austenite''.
In fact, the austenite should be imagined to occupy the region 
$(-\io,0]\times[0,h]$ and to be associated with the elastic energy 
\be 2\p\b \int_{-\io}^0dx\int_0^h |\nabla\psi|^2\;,\label{1.5}\ee
where $\psi$ is periodic in $y$, it decays to zero as $x\to-\io$ and 
satisfies the boundary condition $\psi(0,y)=u(0,y)$. 
Since the elastic energy of the austenite is quadratic, one can perform 
the associated minimization explicitly. This yields 
\be \psi(x,y)=\sum_{k\in\zzzz}\hat u_0(k) e^{2\p i ky/h} e^{2\p |k| x/h}\;,
\label{1.6}\ee
whence
\be 2\p\b \int_{-\io}^0dx\int_0^h |\nabla\psi|^2=4\p^2\b
\sum_{k\in\zzzz}|k||\hat u_0(k)|^2\;.\label{1.7}\ee
Depending on the values of the material parameters $\b,\e$, the minimizers
of (\ref{1.1}) are expected to display different qualitative features.
In particular, in \cite{KM}, on the basis of rigorous upper bounds and 
heuristic lower bounds on the ground state energy of (\ref{1.1}), it was 
conjectured that, if $(\e/L)\ll1$, the minimizers should display periodic 
striped (lamellar) order as long as 
\be \Big(\frac{\e\b}{L}\Big)^{1/2}\ll\Big(\frac\e{L}\Big)^{2/3}\label{1.8}\ee
and asymptotically self-similar branched patterns as long as 
\be \Big(\frac\e{L}\Big)^{2/3}\ll\Big(\frac{\e\b}{L}\Big)^{1/2}\;.\label{1.9}
\ee
Recently, Conti \cite{C06} substantiated this conjecture, by proving that
if $\b\gtrsim\e L/h^2$ then $E_0$, the infimum of (\ref{1.1}) over the 
admissible $u$'s, satisfies upper and lower bounds of the following form:
\be \min\{
c_s\Big(\frac{\e\b}{L}\Big)^{1/2}\,,\; c_b\Big(\frac\e{L}\Big)^{2/3} \}\le 
\frac{E_0}{hL} \le \min\{C_{s}\Big(\frac{\e\b}{L}\Big)^{1/2}\,,\; 
C_b\Big(\frac\e{L}\Big)^{2/3}\}\;,\label{1.11}\ee
for suitable constants $c_s,c_b,C_{s},C_b$. The constants $C_s$ and $C_b$ in 
the r.h.s. are obtained by choosing in the variational upper bound the optimal
periodic striped configurations and the optimal branched configuration, 
respectively.

In the present paper we improve the bounds (\ref{1.11}), by proving 
that, if $\e$ and $\b$ are small and such that (\ref{1.8}) is satisfied, 
i.e., if
\be 0\le \b\ll \Big(\frac{\e}{L}\Big)^{1/3}\ll\frac{h}{L}\;,
\label{1.11z}\ee
then the minimizers display periodic striped order. In particular,
asympotically in the regime (\ref{1.11z}), the constant $c_s$ in the l.h.s. of 
(\ref{1.11}) can be chosen arbitrarily close to $C_s$. Our main result
is summarized in the following theorem. 
\\
\\
{\bf Theorem 1.} {\it If $\e L^2/h^3$ and  
$\b L^{1/3}\e^{-1/3}$ are positive and small enough, any minimizer $u(x,y)$
of (\ref{1.1}) is a one-dimensional periodic sawtooth function, i.e., 
\be u(x,y)=A+w_{M^*}(y-y_0)\;,\label{1.11za}\ee
with $A$ and $y_0$ two real constants, $w_{M}(y):=\int_0^y dz\,
\sign(\sin\frac{\p z M}{h})$ and 
\be M^*={\rm argmin}\{\EE(w_{M})\,:\,M\ {\rm even}\ {\rm integer}\}\;.
\label{1.11zb}\ee 
}
\\
{\bf Remark.} An explicit computation shows that the number
$M^*$ of corner points of the periodic minimizer, as defined in (\ref{1.11zb}),
is $M^*\sim (\b h^2/\e L)^{1/2}\gg1$ for $\b\gg\e L/h^2$, while it is of order 
1 for $\b \lesssim\e L/h^2$.\\ 
\\
In order to prove Theorem 1 we proceed in several steps. First, we  
show that the optimal profile among the one-dimensional (1D) profiles is a sawtooth periodic 
function. This is proved in Section \ref{sec2} and in Appendix \ref{A}, 
by using the reflection positivity method of \cite{GLL1,GLL2,GLL3,GLL4}. 
Next, we show that the minimizers of the full 2D problem are 1D in a subregime 
of (\ref{1.11z}), i.e., for 
\be 0\le \b< (2\p^2h)^{-1/2}\e^{1/2}\;.
\label{1.12z}\ee
The proof of this claim, which is discussed in Section \ref{sec3}, makes use 
both of the lower bound on the energy of 1D configurations of Section 
\ref{sec2} and of a Poincar\'e inequality; the way in which these two bounds 
are combined is the key idea used in the study of the full regime, too.
The proof of Theorem 1 in the full regime (\ref{1.11z}) requires a more 
sophisticated strategy: we first localize the problem in small horizontal 
slices, of vertical size comparable with the optimal period $2h/M^*$, and then 
prove that in each slice $u_x\=0$, by using a combination of Poincar\'e-type 
bounds with a priori estimates on the local 
energy, similar to the one discussed in Section \ref{sec3}.
This is discussed in Section \ref{sec4}. 

\section{Proof of the main result: first step}\label{sec2}
\setcounter{equation}{0}
\renewcommand{\theequation}{\ref{sec2}.\arabic{equation}}

Let us assume that $u_x\=0$ in (\ref{1.1}). In this case $u(x,y)=u(0,y)\=
u_0(y)$
and (\ref{1.1}) reduces to 
\be \EE(u)=\b\int_0^h dy \int_{-\io}^{+\io} dy'\,
\frac{|u_0(y)-\tilde u_0(y')|^2}{|y-y'|^2}+\e L M_0\;,\label{2.1}\ee
where $M_0=N(x=0)$ is the number of jumps of $u_{0}'(y)$. Now, rewrite 
$|y-y'|^{-2}$ as
\be \frac1{|y-y'|^{2}}=\int_0^\io d\a\,\a e^{-\a|y-y'|}\;,
\label{2.2}\ee
so that 
\be \EE(u)=\b\int_0^\io\,d\a\,\a \int_0^h dy \int_{-\io}^{+\io} dy'\,
|u_0(y)-\tilde u_0(y')|^2 e^{-\a|y-y'|}+\e L M_0\;.\label{2.3}\ee
Let us denote by $0\le y_0<y_1<\cdots<y_{M_0-1}<h$ the locations of the 
cusps of $u_0(y)$, and let us define $u_0^{(i)}$, $i=0,\ldots,M_0-1$, 
to be the restrictions of $u_0$ to the intervals $[y_i,y_{i+1}]$. Given 
$u_0^{(i)}$ on $[y_i,y_{i+1}]$, let us extend it to the whole real axis by 
repeated reflections about $y_i$ and $y_{i+1}$; we shall denote the extension 
by $\tilde u_0^{(i)}$. Using the chessboard estimate proved in \cite{GLL3} 
(see Appendix \ref{A} for details)
we find that, for any $\a\in(0,+\io)$, 
\bea &&\int_0^h dy \int_{-\io}^{+\io} dy'\,
|u_0(y)-\tilde u_0(y')|^2 e^{-\a|y-y'|}\ge \label{2.4}\\
&&\hskip1.truecm \ge \sum_{i=0}^{M_0-1}
\int_{y_i}^{y_{i+1}} dy \int_{-\io}^{+\io} dy'\,
|u_0^{(i)}(y)-\tilde u_0^{(i)}(y')|^2 e^{-\a|y-y'|}\;,\nn\eea
which readily implies 
\be \EE(u)\ge \b\sum_{i=0}^{M_0-1}
\int_{y_i}^{y_{i+1}} dy \int_{-\io}^{+\io} dy'\,
\frac{|u_0^{(i)}(y)-\tilde u_0^{(i)}(y')|^2}{
|y-y'|^2}+\e LM_0\;.\label{2.5}\ee
An explicit computation of the integral in 
(\ref{2.5}) gives:
\be \int_{y_i}^{y_{i+1}} dy \int_{-\io}^{+\io} dy'\,
\frac{|u_0^{(i)}(y)-\tilde u_0^{(i)}(y')|^2}{
|y-y'|^2}= \frac{14\,\z(3)}{\p^2} (y_{i+1}-y_i)^2\;,\label{2.6}\ee
where we used that $\sum_{k=1}^\io(2k-1)^{-3}=(7/8) \z(3)$. As a result:
\be \EE(u)\ge \frac{14\,\z(3)}{\p^2} \b\sum_{i=0}^{M_0-1}(y_{i+1}-y_i)^2
+\e LM_0 = \frac{\b c_0 h^2}{M_0}
+\e LM_0+\b c_0\sum_{i=0}^{M_0-1}\Big(h_{i}-\frac{h}{M_0}\Big)^2\;,\label{2.7}\ee
where $c_0=14\,\z(3)/\p^2$ and $h_i=y_{i+1}-y_i$. 
Defining $E_{1D}(M)=\b c_0 h^2/M
+\e LM$ and combining (\ref{2.7}) with the variational bound $\EE(u)\le 
E_{1D}(M^*)$, where $M^*$ is the even integer minimizing $E_{1D}(M)$, 
we find that if $u$ is the minimizer of $\EE(u)$ under the constraint that 
$u_x\=0$,
\be E_{1D}(M^*)\ge \EE(u)\ge E_{1D}(M_0)+\b c_0\sum_{i=0}^{M_0-1}
\Big(h_{i}-\frac{h}{M_0}\Big)^2\;,\label{2.7a}\ee
which implies: (i) $\min\{\EE(u) : u_x\=0\}=E_{1D}(M^*)$; (ii) $M_0=M^*$;
(iii) $h_i=h/M^*$, $\forall i$. Note that even in the 
cases where $E_{1D}(M)$ is minimized by
two distinct values of $M$, $M^*_1$ and $M^*_2$, the only 1D 
minimizers are the simple periodic functions of period $2h/M^*_1$ {\it or}
of period $2h/M^*_2$ (i.e., no function alternating bumps of size $2h/M^*_1$ 
and $2h/M^*_2$ can be a minimizer).

For the purpose of the forthcoming discussion, let us remark that if 
$\b\gg\e L/h^2$, 
\be \Big|M^*-\sqrt{\frac{\b c_0 h^2}{\e L}}\Big|\le 2\label{2.7b}\ee
and 
\be \min\{\EE(u) : u_x\=0\}=E_{1D}(M^*)=hL\,c_s
\sqrt{\frac{\b\e}{L}}\cdot\Big(1+O(\frac{L\e}{h^2\b})\Big)\;,\label{2.7c}\ee
with $c_s=2\sqrt{c_0}$ the constant appearing in (\ref{1.11}).

\section{Proof of the main result: second step}\label{sec3}
\setcounter{equation}{0}
\renewcommand{\theequation}{\ref{sec3}.\arabic{equation}}

The result of the previous section can be restated in the following way:
if $v_M$ is a periodic function 
on $[0,h]$ with $v_M'=\pm1$ and $M$ corners located at $y_i$, $i=1,\ldots,M$, 
then 
\be \b ||v_M||_{H^{1/2}([0,h])}+\e L M\ge E_{1D}(M)+\b c_0\sum_{i=1}^M\Big(h_i-
\frac{h}{M}\Big)^2\;,\label{3.0a}\ee
where $h_i=y_{i+1}-y_i$. In this section we make use of (\ref{3.0a})
and, by combining it with a Poincar\'e inequality, we prove that 
in the regime (\ref{1.12z}) all the minimizers are one-dimensional (and,
therefore, periodic, by the results of Section \ref{sec2}).

Let $M=\min_{x\in [0,L]}N(x)$ and let 
\be \bar x=\inf\{x\in[0,L]\ :\ N(x)=M\}\;.\label{3.0}\ee
Moreover, let $v_M(y)\=u(\bar x,y)$. By the lower semicontinuity of 
$N(x)$ (see the lines following (\ref{1.4a})), $N(\bar x)=M$. We rewrite 
\bea  \EE(u)&=&
\Big[\b||v_M||^2_{H^{1/2}([0,h])}+\e LM\Big] +\b(||u_0||^2_{H^{1/2}([0,h])}-
||v_M||^2_{H^{1/2}([0,h])})+\nn\\
&&+\int_{0}^{L} dx \int_0^h dy\, |u_x|^2 + \e \int_0^L dx \big(N(x)-M\big)\ge\label{3.1}\\
&\ge& \Big[\b||v_M||^2_{H^{1/2}([0,h])}+\e LM\Big] +\b(||u_0||^2_{H^{1/2}([0,h])}-
||v_M||^2_{H^{1/2}([0,h])})+\nn\\
&&+\int_{0}^{\bar x} dx \int_0^h dy\, |u_x|^2 + \e \int_0^{\bar x} dx \big(N(x)-M\big)\;,\nn\eea
where the right hand side of the inequality differs from the left hand side just by the 
upper limits of the two integrals in $dx$, which were set equal to $\bar x$ (in other words, in order
to bound $\EE(u)$ from below, we dropped 
the two positive integrals $\int_{\bar x}^{L} dx \int_0^h dy\, |u_x|^2$
and $\e \int_{\bar x}^L dx \big(N(x)-M\big)$). Now, if $\bar x=0$, then $u$ is 
one-dimensional, $u(x,y)=u_0(y)=v_M(y)$, and we reduce to the discussion in the previous section. Let us then suppose that $\bar x>0$. In this case, 
the first term of the fourth line of Eq.(\ref{3.1}) can be bounded from below by
\be \int_{0}^{\bar x} dx \int_0^h dy\, |u_x|^2\ge \frac1{\bar x}
\int_0^h dy |v_M(y)-u_0(y)|^2
\;.\label{3.2}\ee
The second term of the third line of Eq.(\ref{3.1}) can be rewritten in the form:
\bea \b\big(||u_0||^2_{H^{1/2}([0,h])}-||v_M||^2_{H^{1/2}([0,h])}\big)&=&
\b||u_0-v_M||^2_{H^{1/2}([0,h])}+\nn\\
&+&2\b(v_M,u_0-v_M)_{H^{1/2}([0,h])}
\;,\label{3.3}\eea
where, given two real $h$-periodic functions $f$ and $g$,
\bea (f,g)_{H^{1/2}([0,h])}&=&\int_0^hdy\int_{-\io}^{+\io}dy'\frac{\big(f(y)-
f(y')\big)\big(g(y)-g(y')\big)}{|y-y'|^2}=\nn\\
&=&4\p^2\sum_{k\in\zzzz}|k|\hat f^*(k)\hat g(k)\;.
\label{3.4}\eea
Using Cauchy-Schwarz inequality we find:
\bea \Big| (f,g)_{H^{1/2}([0,h])}\Big|&\le&4\p^2
\sum_{k\in\zzzz}|k|\,|\hat f(k)|\, 
|\hat g(k)|
\le\label{3.5}\\
&\le& 2\p\Big[\frac{4\p^2}h\sum_{k\in\zzzz}|k|^2\,|\hat f(k)|^2\Big]^{1/2}
\cdot \Big[h\sum_{k\in\zzzz}|\hat g(k)|^2\Big]^{1/2} =\nn\\ 
&=&2\p||f'||_{L^2([0,h])}\cdot 
||g||_{L^2([0,h])}\;.\nn\eea
Using (\ref{3.3}), (\ref{3.5}) 
and the fact that $|v_M'|=1$ for a.e. $y$, we find that
\bea \b\big(||u_0||^2_{H^{1/2}([0,h])}-||v_M||^2_{H^{1/2}([0,h])}\big)
&\ge& \b||u_0-v_M||^2_{H^{1/2}([0,h])}-\nn\\
&-&4\p\b h^{1/2} ||u_0-v_M||_{L^2([0,h])}
\;.\label{3.6}\eea
Combining (\ref{3.0a}), (\ref{3.1}), (\ref{3.2}) and (\ref{3.6}), 
and neglecting the positive term $\b||u_0-v_M||^2_{H^{1/2}
([0,h])}$, we get
\bea \EE(u)&&\ge E_{1D}(M)+\b c_0\sum_{i=1}^M\Big(h_i-\frac{h}{M}
\Big)^2-4\p\b h^{1/2} ||u_0-v_M||_{L^2
([0,h])}+\nn\\
&&+ \e \int_0^{\bar x} dx \big(N(x)-M\big)
+\frac1{\bar x}||u_0-v_M||^2_{L^2([0,h])} 
\;.\label{3.7}\eea 
The term $\e\int_0^{\bar x} dx \big(N(x)-M\big)$ is bounded from below 
by $2\e\bar x$ (simply because, by construction,
$N(x)-M\ge 2$ for all $x< \bar x$). Therefore, the last two terms in
the r.h.s. of (\ref{3.7}) are bounded from below by
\bea&&\e \int_0^{\bar x} dx \big(N(x)-M\big)
+\frac1{\bar x}||u_0-v_M||^2_{L^2([0,h])} \ge 2\e\bar x
+\frac1{\bar x}||u_0-v_M||^2_{L^2([0,h])} \nn\\
&&\hskip7.truecm\ge 2\sqrt{2\e}||u_0-v_M||_{L^2([0,h])}\;,\label{3.7a}\eea
which gives us a chance to balance the error term $-4\p\b h^{1/2} 
||u_0-v_M||_{L^2([0,h])}$ in (\ref{3.7}), which is 
linear in $||u_0-v_M||_{L^2([0,h])}$, with 
the sum of the interfacial and the elastic energies.
In fact, by plugging (\ref{3.7a}) into (\ref{3.7}), and neglecting 
a positive term, for any minimizer $u$ we get
\bea && E_{1D}(M^*)\ge \EE(u)\ge E_{1D}(M)+\b c_0\sum_{i=1}^M\Big(h_i-
\frac{h}{M}\Big)^2\label{3.8}\\&&\hskip5.2truecm
+2(\sqrt{2\e}-2\p\b \sqrt{h})||u_0-v_M||_{L^2([0,h])}\;,\nn\eea
where $M^*$ is the even integer minimizing $E_{1D}(M)$.
In the regime (\ref{1.12z}) where $\sqrt{2\e}-2\p\b \sqrt{h}\ge 0$,
Eq.(\ref{3.8}) implies that $u_0\=v_M$, that is,
as observed above, the minimizer is the optimal one-dimensional periodic 
striped state. This concludes the proof of Theorem 1 in the regime 
(\ref{1.12z}).

In the complementary regime 
\be (2\p^2 h)^{-1/2}\e^{1/2}\le \b\ll L^{-1/3}\e^{1/3}\;,\label{4.1}\ee
a similar strategy implies an apriori bound on $M$, which will be
useful in the following. More precisely, by combining 
Eq.(\ref{3.7}) with 
\be\frac1{\bar x}||u_0-v_M
||^2_{L^2([0,h])}-4\p\b\sqrt{h}||u_0-v_M||_{L^2([0,h])}\ge -4\p^2\b^2 h\bar x
\ge -4\p^2\b^2 hL\;,\label{3.8ab}\ee
we find that for any minimizer $u$
\be E_{1D}(M^*)\ge \EE(u)\ge E_{1D}(M)+\b c_0\sum_{i=1}^M\Big(h_i-
\frac{h}{M}\Big)^2-4\p^2\b^2 h L\;.\label{3.8aa}\ee
Recalling that $E_{1D}(M)=\b c_0h^2/M+\e LM$ and the 
fact that $|M^*-\sqrt{\b c_0h^2/\e L}|\le 2$ (see (\ref{2.7b})), from 
(\ref{3.8aa}) we find that
\be \frac{|M-M^*|}{M^*}\le (\const.)\cdot(\b\e^{-1/3}L^{1/3})^{3/4}\ll 1\;.
\label{4.3}\ee
%

\section{Periodicity of the minimizer: the full scaling regime}\label{sec4}
\setcounter{equation}{0}
\renewcommand{\theequation}{\ref{sec4}.\arabic{equation}}

We are now left with proving Theorem 1 in the scaling regime (\ref{4.1}).
In this case the proof is much more elaborate: the rough idea is to apply the 
reasoning of the previous section locally in $y$. We localize 
the functional in horizontal stripes of width $H_j$, comparable with the 
optimal period $2h/M^*\sim \sqrt{\e L/\b}$. In each strip, 
the combination $\sqrt{2\e}-2\p\b \sqrt{h}$ appearing in the right hand side 
of (\ref{3.8}) is replaced by $\sqrt{2\e}- C \b \sqrt{H_j}\ge 
\sqrt{2\e}- C' \b (\e L/\b)^{1/4}$, 
for suitable constants $C,C'$; now,  
the latter expression is $>0$ as long as $\b\ll L^{-1/3}\e^{1/3}$, which will 
allow us to conclude that in every strip the minimizing configuration is 1D. 

In this section, we first discuss how to localize the functional in 
horizontal strips, then we distinguish between ``good'' and ``bad'' 
localization intervals, and finally describe the lower bound on the local 
energy for the different intervals. 

For simplicity, from now on we set $h=L=1$. Here and below
$C,C',\ldots,$ and $c,c',\ldots,$ denote universal constants, 
which might change from line to line.
We assume that $u(x,y)$ is a minimizer, that $\b\ge c\e^{1/2}$, and that 
$\e$ and $\b\e^{-1/3}$ are sufficently small.

\subsection{A localized bound}

Our purpose in this subsection is to derive a local version of the error 
term $-4\p\b h^{1/2}||u_0-v_M||_{L^2([0,h])}$ in Eq.(\ref{3.7}). Let $u_0(y):=u(0,y)$ and 
$u_1(y):=u(1,y)$. Set $F(u)=\int_0^1dx\int_0^1dy
|u_x|^2+\e\int_0^1(N(x)-M)$, denote by $z_i$, $i=1,\ldots,M$, 
the locations of the corners of $u_1$ and by $h_i=z_{i+1}-z_i$ the distances between 
neighboring corners. Note that the number 
of corners of $u_1$ is equal to $M=\min_{x\in[0,1]}N(x)$, because $u$ is a minimizer
and, therefore, $u(x,y)=u(\bar x,y)$ for all $\bar x\le x\le 1$, with $\bar x$ 
defined as in (\ref{3.0}); in fact, the choice $u(x,y)=u(\bar x,y)$ for all $\bar x\le x\le 1$
minimizes the two nonnegative contributions to the energy 
$\int_{\bar x}^{L} dx \int_0^h dy\, |u_x|^2$
and $\e \int_{\bar x}^L dx \big(N(x)-M\big)$, making them precisely zero, see
Eq.(\ref{3.1}).
 
Instead of $v_M$, we now consider a general test function $w(y)$, to be 
specified below, periodic 
on $[0,1]$ and with a number of corners smaller or equal to $M$. 
We denote by $\bar z_i$, $i=0,\ldots M_0-1$, 
the locations of the corners of $w$ (labelled in such a way that 
$0\le \bar z_0<\bar z_1<\cdots<
\bar z_{M_0-1}<1$), and by $\bar h_i=\bar z_{i+1}-\bar 
z_i$ the distances between subsequent corners. In the following, it will be 
useful to imagine that $w$ is associated to a sequence of exactly $M$ corner
points, even in the case that $M_0<M$. These $M$ corner points will be denoted
by $\tilde z_i$, 
$i=0,\ldots,M-1$ and they will have the property that $0\le \tilde z_0\le 
\tilde z_1<\cdots<
\tilde z_{M-1}\le 1$. In the case that $M_0=M$, the sequence of $\tilde z_i$'s 
coincide with the sequence of $\bar z_i$'s; otherwise,
if $M_0<M$, the sequence of the $\tilde z_i$'s will be formed by the 
original sequence of $\bar z_i$'s plus a set of $(M-M_0)/2$ pairs of 
coinciding points. We define 
$\tilde h_i=\tilde z_{i+1}-\tilde z_i$ and note that now, in general, some of 
the $\tilde h_i$'s can be equal to $0$. 

Proceeding as in the previous section, for any minimizer $u$ we get
\bea E_{1D}(M^*)\ge\EE(u)&\ge& \Big[\b||w||^2_{H^{1/2}}+\e M\Big]
+2\b(w,u_0-w)_{H^{1/2}}+F(u) \label{4.8}\\
&\ge & E_{1D}(M)+\b c_0\sum_{i=1}^{M}
\big(\tilde h_i-\frac1M\big)^2+2\b(w,u_0-w)_{H^{1/2}}+F(u)\;.\nn\eea
The first observation is that with the help of the Hilbert transform
we can write (\ref{4.8}) in a more local way. In fact, 
\be (w,u_0-w)_{H^{1/2}}=(\HHH w',u_0-w)_{L^2}\;,\label{4.8a}\ee
with $\HHH$ the Hilbert transform, acting on a periodic function $f$ in the 
following way:
\be (\HHH f)(y)=2\p\sum_{k\neq 0}\frac{-i k}{|k|}\hat f(k) e^{2\p i k y}=
2\p\, P.V.\int_0^1 dy'\cot\p(y-y')\, (f(y')-\bar f)\;,\label{4.9}\ee
where P.V. denotes the Cauchy principal value and $\bar f=\int_0^1 f(y) dy$. 
Combining (\ref{4.8}) and (\ref{4.8a}) we get 
\be E_{1D}(M^*)\ge\EE(u)\ge E_{1D}(M)+\b c_0\sum_{i=1}^{M}
\big(\tilde h_i-\frac1M\big)^2+2\b(\HHH w',u_0-w)_{L^2}+F(u)\;.\label{4.8b}\ee
We now want to bound $2\b(\HHH w',u_0-w)_{L^2([0,h])}$ from below by a sum of 
terms localized in small intervals $I_k\subset [0,h]$, which will 
be the local version of the error term $-4\p\b h^{1/2}||u_0-v_M||_{L^2([0,h])}$
in Eq.(\ref{3.7}). First of all, note 
that, if $\{I_k\}_{k=1,\ldots,M/2}$ is a partition of the unit interval, 
$2\b(\HHH w',u_0-w)_{L^2}$ can be decomposed as
\be 2\b(\HHH w',u_0-w)_{L^2([0,h])}=
2\b \sum_{k=1}^{M/2}\int_{I_k}dy\, \HHH w'(y)\,
(u_0(y)-w(y))\;.\label{4.11a}\ee
In the following we shall choose the partition $\{I_k\}$ in a way depending on 
$u_1$, such that each strip $[0,1]\times I_k$ will typically (i.e., for most 
$k$) contain two or more interfaces of $u$ (as proven by combining the 
definition of $\{I_k\}$ with a priori estimates on $\int_0^1dx\int_{I_k} 
dy\,u_x^2$, see Lemma 1 below). 
Moreover, we shall choose $w$ in a way depending on $\{I_k\}$ and 
on $u_0$, in such a way that every $I_k$ contains at most two corner points of 
$w$ and $\int_{I_k}dy (u_0-w)=0$. Once that $\{I_k\}$ and $w$ are given, every
term in the r.h.s. of (\ref{4.11a}) can be bounded as:
\bea \Big|\int_{I_k}dy\, \HHH w'\,
(u_0-w)\Big|&=&\Big|\int_{I_k}dy (\HHH w'-\overline{\HHH w'})(u_0-w)\Big|
\nn\\
&\le & ||\HHH w'-\overline{\HHH w'}||^2_{L^2(I_k)}||u_0-w||_{L^2(I_k)}\nn\\
&\le & H_k^{1/2} ||\HHH w'||^2_{BMO(I_k)}||u_0-w||_{L^2(I_k)}\;,
\label{4.12}\eea
where $H_k:=|I_k|$, $\overline{\HHH w'}:=|I_k|^{-1}\int_{I_k}dy\,\HHH w'$ 
and the Bounded Mean Oscillation (BMO) seminorm is defined as 
\be ||g||^2_{BMO(I)}=\sup_{(a,b)\subset I}\frac1{|b-a|}\int_a^b dy\, 
|g(y)-g_{(a,b)}|^2\;,\qquad g_{(a,b)}:=\frac1{|b-a|}\int_a^b dy\, g(y)\;.
\label{4.12a}\ee
Now we exploit the fact that the singular
kernel $\cot\p(y-y')$ maps bounded functions into BMO functions \cite{Stein}.
Thus, $||\HHH w'||_{BMO(I_k)}\le C||w'||_{L^\io(I_k)}\le C$, 
uniformly in $w'$ as long as $|w'|\le 1$.  
Therefore, combining (\ref{4.11a}) with (\ref{4.12}), 
we find that there exists a universal constant $\bar c$ such that 
\be 2\b(\HHH w',u_0-w)_{L^2}\ge 
- \bar c\b \sum_{k=1}^{M/2}
H_k^{1/2}||u_0-w||_{L^2(I_k)}\;,\label{4.12c}\ee
which is the desired local version of the error term $-4\p\b h^{1/2}
||u_0-v_M||_{L^2([0,h])}$ in Eq.(\ref{3.7}).

Plugging (\ref{4.12c}) back into (\ref{4.8b}) and using the fact that 
$E_{1D}(M^*)-E_{1D}(M)\le 0$, we find that for any periodic sawtooth function 
$w$ with a number of corners $\le M$,
\be \b c_0\sum_{i=1}^{M}
\big(\tilde h_i-\frac1M\big)^2+F(u)\le \bar c\b \sum_{k=1}^{M/2}
H_k^{1/2}||u_0-w||_{L^2(I_k)}\;,\label{4.11}\ee
which is the main conclusion of this subsection.

\subsection{The choice of the comparison function $w$}

In this subsection we first choose the partition $\{I_k\}$ and the test function $w$ 
to be used in (\ref{4.11}); next, we explain how to use the latter inequality in order
to prove Theorem 1.

Recall that $z_i$, $i=1,\ldots,M$ are the corner
points of $u_1'$. We assume without loss of generality that $u_1'=+1$ in 
$(z_{2k},z_{2k+1})$, $k=1,\ldots,M/2$, and we define $a_k=\frac{z_{2k}+
z_{2k+1}}2$ and $I_k=[a_k,a_{k+1})$, $k=1,\ldots,M/2$ (since we use periodic 
boundary conditions, we shall use the 
convention that $a_0=a_{M/2}$ and $I_0=I_{M/2}$). 
Note that, by construction: (i) $u_1$ 
has exactly two jump points in every interval $I_k$; (ii) the jump points 
are ``well inside'' the intervals $I_k$; (iii) $u_1'(a_k)=+1$; (iv)
$H_k=\frac{h_{2k}}2+h_{2k+1}+\frac{h_{2k+2}}{2}$. 

\begin{figure}[ht]
\hspace{1 cm}
\includegraphics[height=4.cm]{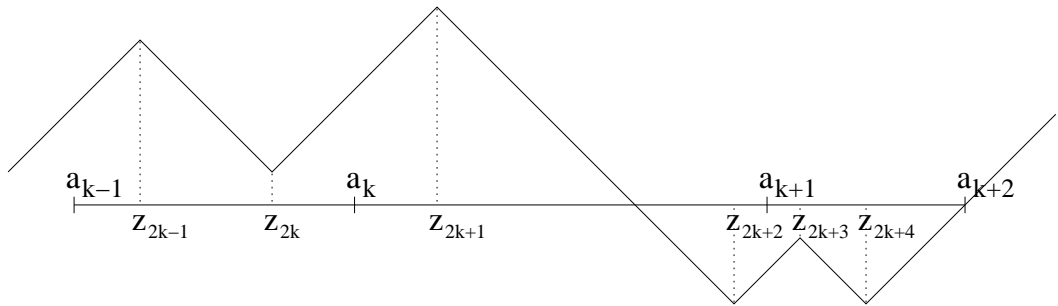}
\caption{The function $u_1$ in the intervals $I_{k-1}=[a_{k-1},a_k)$, 
$I_{k}=[a_{k},a_{k+1})$ and $I_{k+1}=[a_{k+1},a_{k+2})$.}
\end{figure}

Regarding the choice of the test function, 
we choose $w$ to be the 
sawtooth function such that: 
\be {\rm (i)}\ w=u_0\ {\rm on}\ \dpr I_k\,;\qquad {\rm (ii)}\ 
w'=+1\ {\rm on}\ \dpr I_k\,;\qquad
{\rm (iii)}\
\int_{I_k}(w-u_0)=0\;,\ee
for all $k=1,\ldots,M/2$. In every interval $I_k$, $w$ is uniquely 
specified 
by the two corner points $\tilde z_{2k+1},\tilde z_{2k+2}$
chosen in such a way
that: $a_k\le \tilde z_{2k+1}\le \tilde z_{2k+2}\le a_{k+1}$, $w'(y)=+1$ for 
$y\in(a_k,\tilde z_{2k+1})\cup(\tilde z_{2k+2},a_{k+1})$,
$w'(y)=-1$ for $y\in (\tilde z_{2k+1},\tilde z_{2k+2})$ and $\int_{I_k}w=
\int_{I_k}u_0$ (these two corner points are uniquely defined only if $u_0'\not
\=+1$ on $I_k$; if $u_0'\=+1$ on $I_k$, then we set 
$\tilde z_{2k+1}=\tilde z_{2k+2}=\frac{a_{k}+a_{k+1}}{2}$).

\begin{figure}[ht]
\hspace{1 cm}
\includegraphics[height=5.cm]{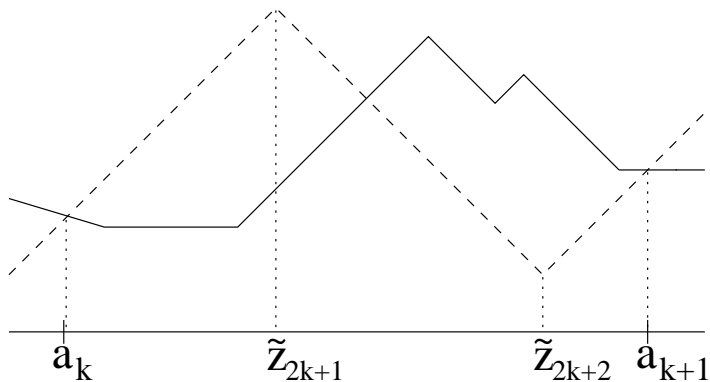}
\caption{The function $u_0$ (full line) and the test function $w$
(dashed line) in the interval $I_{k}=[a_{k},a_{k+1})$.
The function $w$ on $I_k$ and, correspondingly, the locations of 
its corners $\tilde z_{2k+1}$ and $\tilde z_{2k+2}$, are determined by the 
conditions that 
(i) $w=u_0$ on $\dpr I_k$, (ii) $w'=+1$ on $\dpr I_k$, (iii)
$\int_{I_k}(w-u_0)=0$.}
\end{figure}

Note that with the definitions above, $w$ is a sawtooth function with 
$M_0\le M$ corner points, associated to which is a sequence 
$\tilde z_i$, $i=1,\ldots,M$, satisfying the properties described before 
(\ref{4.8}) and $\int_{I_k}(u_0-w)=0$; therefore, $w$ satisfies (\ref{4.12c}).
Let 
\be I^L_k:=\Big[\frac{z_{2k-1}+z_{2k}}2,a_k\Big)\,,\qquad I^R_k:=\Big[a_{k+1},
\frac{z_{2k+3}+z_{2k+4}}2\Big)\,,\qquad I^*_k:=I^L_k\cup I_k \cup I^R_k\;.\label{4.12aa}
\ee
Moreover, let $I^{**}_k:=I_{k-1}\cup I_k\cup I_{k+1}$.
With these definitions, we can rewrite the left hand 
side of (\ref{4.12c}) as $\sum_{k=1}^{M/2}\widetilde F_k$, where
\be \widetilde F_k=\frac{\b c_0}7\sum_{j=2k-2}^{2k+4}\big(\tilde h_{j}-
\frac1{M}\big)^2+\frac13\int_{I^{**}_k}dy\int_0^1 dx\, u_x^2+
\frac{\e}2\int_0^1dx \, \big(N(x)\big|_{I^*_k}-4\big)\;,\label{4.13}\ee
and $N(x)\big|_{I^*_k}$ is the number of corner points of $u(x,\cdot)$ in 
$I^*_k$. In the following we shall denote by 
$\widetilde F_k^{(0)}$ the first term in the r.h.s. of (\ref{4.13}),
by $\widetilde F_k^{(1)}$ the second term and by $\widetilde F_k^{(2)}$ the 
third term. Using these definitions, (\ref{4.11}) can be rewritten as
\be\sum_{k=1}^{M/2}\big(\widetilde F_k-\bar c\b
H_k^{1/2}||u_0-w||_{L^2(I_k)}\big)\le 0\;.\label{4.14}\ee
Our next goal is to derive a lower bound on the l.h.s. of (\ref{4.14})
of the form 
\be \sum_{k=1}^{M/2}\big(\widetilde F_k-\bar c\b H_k^{1/2}
||u_0-w||_{L^2(I_k)}\big)\ge \frac12\sum_{k=1}^{M/2}\big(\widetilde F_k^{(0)}+
\widetilde F_k^{(1)}\big)+\Big(1-(\b\e^{-1/3})^\a\Big)
\sum_{k=1}^{M/2}\widetilde F_k^{(2)}\;,\label{4.14a}\ee
for a suitable $\a>0$. Plugging (\ref{4.14a}) into (\ref{4.14}) gives 
\be \frac{\b c_0}2\sum_{i=1}^{M}\big(\tilde h_{i}-\frac1{M}\big)^2+\frac12
\int_0^1 dy\int_0^1 dx\, u_x^2+\e\Big(1-(\b\e^{-1/3})^\a\Big)\int_0^1dx\,
\big(N(x)-M\big)\le 0\;,\label{4.14b}\ee
which implies that $u_x\=0$ and $N(x)\=M$, and concludes the proof of 
Theorem 1.

The rest of the paper will be devoted to the proof of (\ref{4.14a}).
In order to get bounds from above on $H_k^{1/2}||u_0-w||_{L^2(I_k)}$,
it will be convenient to distinguish between ``good'' and ``bad'' intervals, 
and to proceed in different ways, depending on the nature of the interval 
$I_k$.

\subsection{Classification of the good and bad intervals}

We shall say that\begin{itemize}
\item $I_k$ is ``good'' (of type 1) if $\max_{k-1\le i\le k+1}H_i\le 6/M$, 
$\widetilde F^{(1)}_k\le \h/M^3$, and $\min_{2k-1\le j\le 2k+3}
h_j\ge \k/M$, for suitable constants $\h,\k$, to be conveniently fixed
below.
\end{itemize} 
Note that if $u(x,y)$ is the periodic sawtooth function in (\ref{1.11za}), 
as we hope to prove, then all the 
intervals $I_k$ are good. Conversely, if $I_k$ is good, then $u(x,y)|_{y\in 
I^*_k}$ is in 
some sense close to the optimal 1D configuration. More precisely, if $I_k$ is 
good, then its length is of the same order as $2/M$; moreover, the corners of 
$u_1|_{I^*_k}$ are well separated, on the same scale, and $u_1|_{I^{**}_k}$ is 
close to $u_0|_{I^{**}_k}$ in $L^2$, on the natural scale: 
in fact, by the Poincar\'e inequality, $||u_0-u_1||^2_{L^2(I^{**}_k)}\le 
3\widetilde F^{(1)}_k\le 3\h/M^3$. 

The ``bad'' intervals will
be further classified in three different types; we shall say that:
\begin{itemize}
\item $I_k$ is of type 2 if $\max_{k-1\le i\le k+1}H_i\le 6/M$, 
$\widetilde F^{(1)}_k\le \h/M^3$, and $\min_{2k-1\le j\le 2k+3}
h_j< \k/M$;
\item $I_k$ is of type 3 if $\max_{k-1\le i\le k+1}H_i\le 6/M$ and 
$\widetilde F^{(1)}_k> \h/M^3$;
\item $I_k$ is of type 4 if $\max_{k-1\le i\le k+1}H_i> 6/M$.
\end{itemize}
We denote by $\III_q$, $q=1,\ldots,4$, the set of intervals of type $q$;
note that $\cup_q\III_q=\cup_k\{I_k\}$. 
In the following we describe how to obtain upper bounds on 
$\bar c\b\sum_{k: I_k\in\III_q} H_k^{1/2}||u_0-w||_{L^2(I_k)}$ 
of the form (\ref{4.14a}), separately for $q=1,2,3,4$.
Here and below we denote by $c,c',C,C',\ldots,$ universal constants 
independent of $\h,\k$.

\subsection{The lower bound: the good intervals}

For intervals of type 1, the key estimates to be proven are the following.

{\bf Lemma 1} {\it Let $I_k$ be an interval of type 1. If $\h \k^{-3}$ 
is small enough, then
$N(x)\Big|_{I_k}\ge 2$, $N(x)\Big|_{I^L_k}\ge 1$ and 
$N(x)\Big|_{I^R_k}\ge 1$, $\forall x\in[0,1]$.}

{\bf Lemma 2} {\it Let $I_k$ be an interval of type 1. Let us 
define $\bar x_k=\inf_{x\in[0,1]}\{x\ :\ N(x)|_{I_k^*}\le 4\}$ and $\bar u(y)\=
u(\bar x_k,y)$. If $\k$ and $\h\k^{-3}$ are small enough,
then there exists a constant $C$ independent of $\h,\k$ such that 
\be||u_0-w||_{L^2(I_k)}\le C \k^{-5/2}
||u_0-\bar u||_{L^2(I_k^{**})}\;.\label{l2}\ee}
We first show that Lemma 1 and 2 imply
the desired bound,  
\be \bar c\b H_k^{1/2}||u_0-w||_{L^2(I_k)}\le c \k^{-5/2}
(\b\e^{-1/3})^{3/4}\,\widetilde F_k\;.\label{4.15}\ee
Note that (\ref{4.15}) implies (\ref{4.14a}) for all $\a<3/4$ 
(because $\widetilde F^{(2)}_k\ge 0$ for intervals of type 1). 
The strategy to prove (\ref{4.15}) from Lemma 1 and 2 is the same followed in 
Section \ref{sec3} to prove (\ref{3.7a}): we use an interpolation 
between the interfacial energy and the elastic energy to get a lower
bound for $\widetilde F_k$, which is linear in $||u_0-w||_{L^2(I_k)}$. 
In fact, if $\bar x_k=0$, then by definition $\bar u=u_0$ and, by Lemma 2,
$u_0\=w$ on $I_k$, in which case (\ref{4.15}) is obvious. 
If $\bar x_k>0$, then, by Lemma 1, $\widetilde F_k\ge\frac13 
\int_0^{\bar x_k}dx \int_{I_k^{**}}u_x^2+\frac{\e}2\int_0^{\bar x_k}
\big(N(x)\big|_{I_k^*}-4\big)$. Using the Poincar\'e inequality and
the fact that, by definition of $\bar x_k$, $N(x)\big|_{I_k^*}-4\ge 1$ if 
$0\le x<\bar x_k$, we find: 
\bea \widetilde F_k\ge \frac1{3\bar x_k}||u_0-\bar u
||^2_{L^2(I_k^{**})}+\frac{\e}2\bar x_k&\ge& \Big(\frac{2\e}3\Big)^{1/2}
||u_0-\bar u||_{L^2(I_k^{**})}\ge\nn\\
&\ge& c \e^{1/2}\k^{5/2}
||u_0-w||_{L^2(I_k)}\;,\label{4.16}\eea
where in the last inequality we used Lemma 2. Using (\ref{4.16}) and the fact
that for type 1 intervals $H_k\le 6/M\le c \e^{1/2}
\b^{-1/2}$ (see (\ref{2.7b}) and (\ref{4.3})), we find (\ref{4.15}). 
Let us now prove Lemma 1 and 2.\\

{\cs Proof of Lemma 1.} Let us start by showing that $N(x)|_{I_k}\ge 2$.
Let us assume by contradiction
that there exists $x^*$ such that $N(x^*)|_{I_k}<2$. Let $v(y)\=u(x^*,y)$ 
and let us consider the intervals $J_{k,1}=(z_{2k+1}-\frac{\k}{4M},z_{2k+1}+
\frac{\k}{4M})$ and $J_{k,2}=(z_{2k+2}-\frac{\k}{4M},z_{2k+2}+
\frac{\k}{4M})$. Note that by the definition of type 1 intervals, $J_{k,1}$ and 
$J_{k,2}$ are disjoint and both contained in $I_k$. Since $v(y)$ has less than 
two corner points in $I_k$, then $v(y)$ has no corner points in at least 
one of the two intervals $J_{k,1}$ and $J_{k,2}$, say in $J_{k,1}$. Now, 
$3\widetilde F^{(1)}_k\ge \int_{x^*}^1dx \int_{I_k}dy\, u_x^2\ge 
||u_1-v||^2_{L_2(J_{k,1})}$. 
Using that $v$ has no corners in $J_{k,1}$, one finds that 
$3\widetilde F^{(1)}_k\ge||u_1-v||^2_{L_2(J_{k,1})}\ge c\k^3/M^3$, 
a contradiction if $\h\k^{-3}$ is sufficiently small. The proof that 
$N(x)|_{I^{L,R}_{k}}\ge 1$ is completely analogous. 
This proves Lemma 1. Moreover, it proves that $u(x,\cdot)$
has at least one corner in each of the intervals $J_{k-1,2}$,
$J_{k,1}$, $J_{k,2}$, $J_{k+1,1}$, $\forall x\in[0,1]$.\qed

\vspace{.5truecm}
{\cs Proof of Lemma 2.} By the definition of $\bar x_k$ and by the result of Lemma 1, 
$\bar u(y)=u(\bar x_k,y)$ has exactly 1 corner point in $I^{L}_{k}$
(located in $J_{k-1,2}$), exactly 1 corner point in $I^{R}_{k}$ (located in 
$J_{k+1,1}$) and exactly 2 corner points in $I_{k}$ (one located in $J_{k,1}$
and one in $J_{k,2}$). We 
shall denote by $z^*_j$, $j=0,1,2,3$, these corner points (with $z^*_0<z^*_{1}
<z^*_2<z^*_3$). Moreover, $\bar u'(y)=+1$ if $y\in(z^*_0,z^*_1)\cup (z^*_2,
z^*_3)$ and $\bar u'(y)=-1$ if $y\in(z^*_1,z^*_2)$. By the definition of $J_{k,1}$ 
and $J_{k,2}$, we have that
$z^*_2-z^*_1\ge \k/(2M)$ and $\min\{a_k-z^*_0,z^*_1-a_k, a_{k+1}-z^*_2,
z^*_3-a_{k+1}\}\ge \k/(4M)$.

Let $\d:= M^{3/2}||u_0-\bar u||_{L^2(I^{**}_k)}$. 
If $\d>\d_0$, with $\d_0=\bar c_0 \k^{5/2}$, then (\ref{l2}) is 
proved; in fact, in this case, since $u_0=w$ on $\dpr I_k$ and $|(u_0-w)'|\le 2$,
$||u_0-w||_{L^2(I_k)}
\le c' H_k^{3/2}\le c''M^{-3/2}\le (c''/\d_0)
||u_0-\bar u||_{L^2(I^{**}_k)}=c'''\k^{-5/2}||u_0-\bar u||_{L^2(I^{**}_k)}$,
which is the desired estimate.

Let then $\d\le \d_0$ and let us note that $||w-\bar u||_{L^\io(\partial I_k)}=
||u_0-\bar u||_{L^\io(\partial I_k)}\le 4\k^{-1/2}\d M^{-1}$.
Indeed, if, by contradiction, $u_0(a_k)-\bar u(a_k)> 4\k^{-1/2}\d M^{-1}$, then
$u_0(x)-\bar u(x)=u_0(a_k)-\bar u(a_k)+\int_{a_k}^x(u_0'-1)(y) dy\ge 
u_0(a_k)-\bar u(a_k)>4\k^{-1/2}\d M^{-1}$, for all $x\in(z_0^*, a_k)$ (here we
used that $\bar u'=1$ in $(z_0^*,z_1^*)$ and $|u_0'|\le 1$); 
similarly, if $\bar u(a_k)-u_0(a_k)> 4\k^{-1/2}\d M^{-1}$, then
$\bar u(x)-u_0(x)\ge \bar u(a_k)-u_0(a_k)>4\k^{-1/2}\d M^{-1}$, 
for all $x\in(a_k, z_1^*)$; in both cases, using the fact 
that $\min\{a_k-z^*_0,z^*_1-a_k\}\ge \k/(4M)$, we would find 
$\d M^{-3/2}\=||u_0-\bar u||_{L^2(I^{**}_k)}>2\d M^{-3/2}$, a contradiction. 
Now, let $g=w-\bar u$ and let $g^*=
g(y^*)$, with $y^*\in I_k$, such that $|g(y^*)|=||g||_{L^\io(I_k)}$. We 
want to prove that if $\k$ is sufficiently small, then $|g^*|\le
\k^{-5/2}\d M^{-1}$; if this is the case, 
then $||u_0-w||_{L^2(I_k)}\le \d M^{-3/2}+||g||_{L^2(I_k)}\le \d M^{-3/2}+
\sqrt{6} M^{-1/2}\k^{-5/2}\d M^{-1}=(1+\sqrt{6}
\k^{-5/2})||u_0-\bar u||_{L^2(I^{**}_k)}$,
which is the desired bound. 

Let us then assume by contradiction that $|g^*|> \k^{-5/2}\d M^{-1}$. 
Note that by construction $g$ has the following properties:
\begin{enumerate}
\item $||g||_{L^\io(\partial I_k)}\le 4\k^{-1/2}\d M^{-1}$;
\item there exist $y_1,y_2,y_3,y_4$ such that: (i)
$a_k\le y_1\le y_2\le y_3\le y_4\le a_{k+1}$; (ii) 
$g'(y)=0$ for $y\in(a_k,y_1)\cup(y_2,y_3)\cup(y_4,a_{k+1})$,
$g'(y)=m$ for $y\in(y_1,y_2)$ and $g'(y)=-m$ for $y\in(y_3,y_4)$, with
$|m|=2$;
\item if we define $\D_1=y_2-y_1$, $\D_2=y_3-y_2$ and $\D_3=y_4-y_3$, 
then $\D_1+\D_2+\D_3\ge z_2^*-z_1^*\ge \k/(2M)$. 
\end{enumerate}

\begin{figure}[ht]
\hspace{1 cm}
\includegraphics[height=5.5cm]{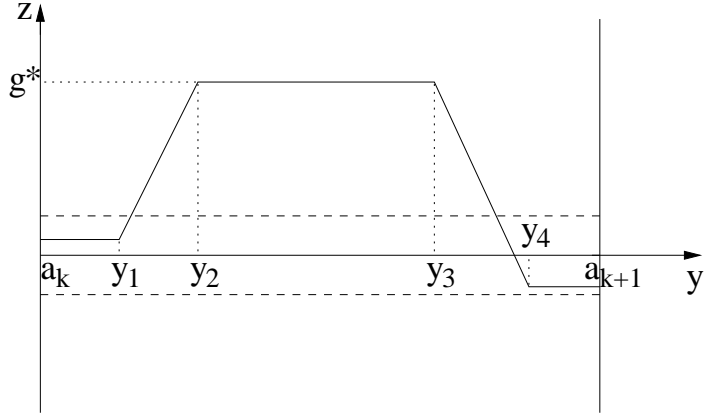}
\caption{The function $z=g(y)$ in the interval $I_{k}=[a_{k},a_{k+1})$.
The two horizontal dashed lines are $z=\pm 4\k^{-1/2}\d M^{-1}$. 
Since $||g||_{L^\io(\partial I_k)}
\le 4\k^{-1/2}\d M^{-1}$, the two horizontal portions of the graph of $g$ in 
the intervals $(a_k,y_1)$ and $(y_4,a_{k+1})$ stay inside the strip
$-4\k^{-1/2}\d M^{-1}\le z\le 4\k^{-1/2}\d M^{-1}$.}\label{fig3}
\end{figure}

Let us assume without loss of generality that $m=+2$, so that 
$g^*=\max_{I_k}g>\k^{-5/2}\d M^{-1}$ and 
$(g^*-4\k^{-1/2}\d M^{-1})/2\le \D_i\le (g^*+4\k^{-1/2}\d M^{-1})/2$, both for
$i=1$ and $i=3$. 
Now, if $\D_2\le \k/(4M)$, then $\D_1+\D_3\ge 
\k/(4M)$ and $\D_1^2+\D_3^2\ge \k^2/(32 M^2)$. On the other hand, 
using that $\int_{I_k}(u_0-\bar u)=\int_{I_k}g$, we find:
\be \d M^{-3/2}=||u_0-\bar u||_{L^2(I^{**}_k)}\ge H_k^{-1/2}\Big|
\int_{I_k}(u_0-\bar u)\Big|=H_k^{-1/2}\Big|
\int_{I_k}g\Big|\ge  \frac{M^{1/2}}{\sqrt{6}} \Big|
\int_{I_k}g\Big|\;.\label{4.17aa}\ee
Now, denoting by $\widetilde y_1$ and $\widetilde y_4$ the two points $\widetilde y_1=
y_2-\frac{g^*-4\k^{-1/2}\d M^{-1}}{2}$ and
$\widetilde y_4=y_3+\frac{g^*-4\k^{-1/2}\d M^{-1}}{2}$ such that 
$g(\widetilde y_1)=g(\widetilde y_4)=+4\k^{-1/2}\d M^{-1}$ (see Fig.\ref{fig3}), 
we can bound $|\int_{I_k}g|$
from below as $|\int_{I_k}g|\ge \int_{\tilde y_1}^{y_2}[g^*-2(y_2-y)]dy+
\int_{y_3}^{\tilde y_4}[g^*-2(y-y_3)]dy-4\k^{-1/2}\d M^{-1}[(\widetilde y_1-a_k)+(a_{k+1}-
\widetilde y_4)]$, which implies
\bea\d M^{-3/2}&\ge&  \frac{M^{1/2}}{\sqrt{6}}
\Big[\frac{(g^*)^2-16\k^{-1}\d^2 M^{-2}}2
-C'\k^{-1/2}\d M^{-2}\Big]\ge\nn\\
&\ge & \frac{M^{1/2}}{\sqrt{6}}\Big[\frac{g^*-4\k^{-1/2}\d M^{-1}}{g^*+
4\k^{-1/2}\d M^{-1}}
(\D_1^2+\D_3^2)-C'\k^{-1/2}\d M^{-2}\Big]>\nn\\
&>&  c' M^{1/2}\Big[\frac{\k^2}{M^2}-C''\k^{-1/2}\d M^{-2}\Big]
\;,\label{4.18}\eea
where in the last inequality we used $g^*>\k^{-5/2}\d M^{-1}$ and the fact 
that $\k$ is sufficiently small. Eq.(\ref{4.18}) implies 
$$\bar c_0\k^{5/2}=\d_0\ge \d> c''\k^{5/2}\;,$$
a contradiction if $\bar c_0$ is chosen small enough. Finally, if $\D_2> 
\k/(4M)$, then 
\bea \d M^{-3/2}&\ge& \frac{M^{1/2}}{\sqrt{6}}\Big|
\int_{I_k}g\Big|\ge \frac{M^{1/2}}{\sqrt{6}}\Big[
\D_2g^*-C'\k^{-1/2}\d M^{-2}\Big] \nn\\
&>&c'\k^{-3/2}\d M^{-3/2}-C''\k^{-1/2}\d M^{-3/2}\;,
\label{4.19}\eea
which leads to a contradiction if $\k$ is sufficiently small.
This concludes the proof of Lemma 2.\qed

\subsection{The lower bound: the bad intervals}

For intervals of type 2, 3 and 4, the key estimate that we shall use is 
the following.

{\bf Lemma 3} {\it Let $I_k$ be an interval of any type. There exists a 
constant $C$ independent of $\h,\k$ such that 
\be ||u_0-w||_{L^2(I_k)}\le C H_k ||u_0-u_1||^{1/3}_{L^2(I_k)}\;.\label{l3}\ee}
{\cs Proof of Lemma 3.} 
First of all, note that (\ref{l3}) is invariant under the rescaling 
$I_k\to\tilde I^{(\ell)}_k
=[\ell a_k,\ell a_{k+1})$ combined with 
$u(y)\to\tilde u^{(\ell)}(y)=\ell u(y/\ell)$;
therefore, we can freely assume that $H_k=1$ and we denote by $I=[0,1)$ the 
corresponding rescaled (unit) interval. Let $y^*$ be such that 
$|(u_1-u_0)(y^*)|=||u_1-u_0||_{L^\io(I)}$; 
using that $|(u_1-u_0)'|\le 2$, we find that 
$||u_1-u_0||_{L^\io(I)}\le |(u_1-u_0)(y)|+2|y-y^*|$, $\forall y\in I$. 
Without loss of generality, we can assume that $y^*$ is in the left half
of $I$, in which case, for any $0\le \d\le 1/2$,
\be ||u_1-u_0||_{L^\io(I)}\le\int_{y^*}^{y^*+\d}\frac{dy}{\d}|(u_1-u_0)(y)|+
\d\le\d^{-1/2}||u_1-u_0||_{L^2(I)}+\d\;.\label{4.24aa}\ee
Now, if $||u_1-u_0||_{L^2(I)}\ge 2^{-3/2}$, then (\ref{l3}) is trivial: 
in fact $||u_0-w||_{L^2(I)}\le 1/\sqrt3$,
simply because $|(u_0-w)(y)|\le 2\min\{y,1-y\}$, and, therefore,   
$||u_0-w||_{L^2(I)}\le \sqrt{2/3}\,||u_1-u_0||_{L^2(I)}^{1/3}$,
which is the desired estimate. 

Let us then suppose that 
$||u_1-u_0||_{L^2(I)}< 2^{-3/2}$. In this case, choosing 
$\d=||u_1-u_0||_{L^2(I)}^{2/3}$ in (\ref{4.24aa}), we find that $\t:=||u_1-u_0
||_{L^\io(I)}\le 2||u_1-u_0||_{L^2(I)}^{2/3}<1$. 
Let us now define, in analogy 
with the proof of (\ref{4.18})-(\ref{4.19}), $g=w-u_1$, and let $g^*=g(y^*)$,
with $y^*\in[0,1]$ such that $|g(y^*)|=||g||_{L^\io(I)}$. Note that 
by construction $g$ has the following properties:
there exist $y_1,y_2,y_3,y_4$ such that $0\le y_1\le y_2\le y_3\le y_4\le 1$
and $g'(y)=0$ for $y\in(0,y_1)\cup(y_2,y_3)\cup(y_4,1)$,
$g'(y)=m$ for $y\in(y_1,y_2)$ and $g'(y)=-m$ for $y\in(y_3,y_4)$, with
$|m|=2$. We also define $\D_1=y_2-y_1$, $\D_2=y_3-y_2$ and $\D_3=y_4-y_3$.
Let us distinguish
two more subcases.\begin{enumerate}
\item $|g^*|< 9\t$.
In this case, $||u_0-w||_{L^2(I)}\le ||u_0-w||_{L^\io(I)}
\le 10||u_1-u_0||_{L^\io(I)}
\le 20||u_1-u_0||_{L^2(I)}^{2/3}\le 2^{1/2}\cdot10
||u_1-u_0||_{L^2(I)}^{1/3}$, which is the desired bound.
\item $|g^*|\ge 9\t$. In this case, proceeding as in the proof
of (\ref{4.18})-(\ref{4.19}), we find:
\bea \t\ge|\int_I(u_1-u_0)|&=&|\int_I(u_1-w)|\ge 
\frac{1}{2}(\D_1^2+\D_3^2)+|g^*|\D_2-\t\nn\\
&\ge &\frac{1}{4}(\D_1+\D_3)^2+|g^*|\D_2-\t
\;.\label{C.2}\eea
If $\D_1+\D_3\ge 3\sqrt{\t}$ or $\D_2\ge 1/4$, then (\ref{C.2}) implies that 
$2\t\ge 9\t/4$, which is a contradiction. 
Therefore, $\D_1+\D_3< 3\sqrt{\t}$ and $\D_2< 1/4$; using that $(8/9)|g^*|\le
|g^*|-\t\le \D_1+\D_3$, we get $|g^*|\le (27/8)\sqrt{\t}$.
In conclusion, $||u_0-w||_{L^2(I)}\le ||u_0-w||_{L^\io(I)}
\le \t+(27/8)\sqrt{\t}< (35/8)\sqrt{\t}\le (35\sqrt2/8)
||u_1-u_0||_{L^2(I)}^{1/3}$, which is the desired estimate.\qed
\end{enumerate}

Let us now show how to use Lemma 3 in order to get a bound from above
on $\sum_{k: I_k\in\III_q}\big(\widetilde F_k-\bar c\b H_k^{1/2}
||u_0-w||_{L^2(I_k)}\big)$, separately for $q=2,3,4$.

\subsubsection{Intervals of type 2}

In this case, the key remark is that, if $\k$ and $\h\k^{-3}$ are small enough,
then necessarily 
%
\be \min_{2k-1\le j\le 2k+3}\tilde h_{j}\le 
\frac1{2M}\;.\label{4.20}\ee
Let us prove this fact. If $\min\{H_{k-1},H_k,H_{k+1}\}<1/(2M)$ 
the claim is obvious, so let us assume that $\min\{H_{k-1},H_k,H_{k+1}\}
\ge 1/(2M)$. Let us first consider the case that $h_{2k^*+1}
:=\min\{h_{2k-1},h_{2k+1},h_{2k+3}\}<\k/M$. In this case, using that 
$h_{2k^*+1}=H_{k^*}/2-[u_1(a_{k^*+1})
-u_1(a_{k^*})]/2$ and $\tilde h_{2k^*+1}=H_{k^*}/2-[u_0(a_{k^*+1})
-u_0(a_{k^*})]/2$, we find that $|h_{2k^*+1}-\tilde h_{2k^*+1}|\le 
2||u_1-u_0||_{L^\io(I_{k^*})}$. On the other hand, if $y^*$ is such that 
$|(u_1-u_0)(y^*)|=||u_1-u_0||_{L^\io(I_{k^*})}$, 
using that $|(u_1-u_0)'|\le 2$, we have that 
$||u_1-u_0||_{L^\io(I_{k^*})}\le |(u_1-u_0)(y)|+2|y-y^*|$, $\forall y\in 
I_{k^*}$. Proceeding as in the proof of (\ref{4.24aa}), we find that for any 
$\d\le H_{k^*}/2$,
\be ||u_1-u_0||_{L^\io(I_{k^*})}\le\d^{-1/2}||u_1-u_0||_{L^2(I_{k^*})}+\d\;.
\label{4.24a}\ee
Choosing $\d=||u_1-u_0||_{L^2(I_{k^*})}^{2/3}$, which, by Poincar\'e 
inequality, is smaller than $[3\widetilde F^{(1)}_{k^*}]^{1/3}\le 
(3\h)^{1/3}M^{-1}$ (which is in turn smaller than $H_{k^*}/2$ for $\h$ small 
enough), we find:
\bea &&|h_{2{k^*}+1}-\tilde h_{2{k^*}+1}|\le 
2||u_1-u_0||_{L^\io(I_{k^*})}\le 4||u_1-u_0||_{L^2(I_{k^*})}^{2/3}\le
4(3\h)^{1/3}M^{-1}\ \ \Rightarrow\nn\\
&&\quad \Rightarrow\quad \tilde h_{2{k^*}+1}\le \frac{\k+4
(3\h)^{1/3}}{M}\le \frac1{2M}\;,\label{4.24b}\eea
where in the last inequality we assumed that $\k$ and $\h$ are small enough.

By definition of type 2 intervals, we are left with the case that 
$\min\{h_{2k},h_{2k+2}\}<\k/M$. 
Without loss of generality, we can assume that
$h_{2k}<\k/M$  and $\min\{h_{2k-1},h_{2k+1},h_{2k+3}\}\ge \k/M$;
by contradiction, we assume that $\tilde h_{2k}\ge 1/(2M)$, so that 
$\max\{\tilde z_{2k+1}-a_{k}, a_k-\tilde z_{2k}\}\ge 1/(4M)$,  
say $\tilde z_{2k+1}-a_{k}\ge 1/(4M)$.

By (\ref{4.24a}), $\t:=||u_1-u_0||_{L^\io(I_k^{**})}\le \d^{-1/2} 
||u_1-u_0||_{L^2(I_k^{**})}+\d$, so that, choosing $\d=||u_1-u_0
||_{L^2(I_k^{**})}^{2/3}$, we get $||u_1-u_0||_{L^\io(I_k^{**})}\le
2\d\le 2(3\h)^{1/3}M^{-1}$, by Poincar\`e (see the lines following 
(\ref{4.24a})). Proceeding in a way analogous 
to the proof of (\ref{4.18})-(\ref{4.19}), we define $g=w-u_1$, so that:
\be \frac{2(3\h)^{1/3}}M \ge \t \ge H_k^{-1}\Big|\int_{I_k}
(u_0-u_1)\Big|\ge \frac{M}{6}\big|\int_{I_k}g\big|\;.\label{4.24c}\ee
Recall the assumptions on $h_i$ and $\tilde h_{2k}$:  
$h_{2k}<\k/M$, $\min\{h_{2k-1},h_{2k+1},h_{2k+3}\}\ge \k/M$, and 
$\tilde z_{2k+1}-a_{k}\ge 1/(4M)$. Therefore, $g$ has the following properties:
there exist $y_1,y_2,y_3,y_4$ such that $a_k\le y_1\le y_2\le y_3\le y_4\le 
a_{k+1}$ and $g'(y)=0$ for $y\in(0,y_1)\cup(y_2,y_3)\cup(y_4,1)$,
$g'(y)=m$ for $y\in(y_1,y_2)$ and $g'(y)=-m$ for $y\in(y_3,y_4)$, with
$|m|=2$; moreover, if $\D_1:=y_2-y_1$ and 
$\D_2=y_3-y_2$, then $\D_1\ge \k/M$ and $\D_1+\D_2\ge (1-2\k)/(4M)$. If 
$\k$ and $\h\k^{-3}$ are sufficiently small, by proceeding as in the proof of 
(\ref{4.18}) and (\ref{4.19}), we can bound (\ref{4.24c}) from below by
\be \frac{2(3\h)^{1/3}}M\ge \t\ge c M(\D_1^2+\D_1\D_2-c'\t)\ge \frac{c''\k}M\;,
\label{4.24d}\ee
which is a contradiction. If $h_{2k}<\k/M$, $\min\{h_{2k-1},h_{2k+1},h_{2k+3}
\}\ge \k/M$, and $a_k-\tilde z_{2k}\ge 1/(4M)$, one can proceed in a completely
analogous way, by replacing $I_k$ by $I_{k-1}$ in (\ref{4.24c}).
This concludes the proof of (\ref{4.20}).
\\

Once that (\ref{4.20}) is proved, we find that 
\be \frac{\b c_0}7\sum_{j=2k-2}^{2k+4}\big(\tilde h_{j}-\frac1{M}
\big)^2\ge \frac{c\b}{M^2}\label{4.21}\ee
and, as a consequence, defining $\widetilde F^{(01)}_k:=\widetilde F^{(0)}_k+
\widetilde F^{(1)}_k$, 
\be \widetilde F_2:=
\sum_{k: I_k\in\III_2}\widetilde F_k^{(01)}\ge \frac{c\b}{M^2}\NN_2
\quad \Rightarrow\quad \NN_2\le c^{-1}\b^{-1} M^2\widetilde F_2\;,
\label{4.22}\ee
where $\NN_2=|\III_2|$ is the number of intervals of type 2. Now, by Lemma 3
and the fact that $||u_0-u_1||_{L^2(I_k)}^2\le \int_0^1dx \int_{I_k}
dy\, u_x^2\le 3\widetilde F_k^{(01)}$, we have that 
\bea \bar c\b \sum_{k: I_k\in\III_2} H_k^{1/2}||u_0-w||_{L^2(I_k)}
&\le& c'\b \sum_{k: I_k\in\III_2} H_k^{3/2}||u_0-u_1||_{L^2(I_k)}^{1/3}
\le\nn\\
&\le& c''\b \sum_{k: I_k\in\III_2} M^{-3/2}[\widetilde 
F_k^{(01)}]^{1/6}\;.
\label{4.23}\eea
Using Minkowski's inequality, we find:
\be  \sum_{k: I_k\in\III_2} [\widetilde 
F_k^{(01)}]^{1/6}\le \Big[\sum_{k: I_k\in\III_2} \widetilde 
F_k^{(01)}\Big]^{1/6}
\Big[\sum_{k: I_k\in\III_2} 1\Big]^{5/6}\le c'\widetilde F_2^{1/6}
\Big[M^2\b^{-1}\widetilde F_2\Big]^{5/6}\;.\label{4.24}\ee
Combining (\ref{4.23}) and (\ref{4.24}), we find that 
\be \bar c\b \sum_{k: I_k\in\III_2} H_k^{1/2}||u_0-w||_{L^2(I_k)}\le
c''\b^{1/6} M^{1/6} \widetilde F_2\le c''' (\b\e^{-1/3})^{1/4}
\widetilde F_2\;,\label{4.240}\ee
where in the last inequality we used that $M\le c (\b/\e)^{1/2}$, see 
Eq.(\ref{2.7b}). By using (\ref{4.240}), defining $\s:=\b\e^{-1/3}$ and for 
any $\a>0$, we get
\bea  \sum_{k: I_k\in\III_2} \big(\widetilde F_k- \bar c\b 
H_k^{1/2}||u_0-w||_{L^2(I_k)}\big)&\ge& \frac12\sum_{k: I_k\in\III_2} 
\widetilde F_k^{(01)}+\big(1-\s^\a\big)\sum_{k: I_k\in\III_2} 
\widetilde F^{(2)}_k+\nn\\
&+&\sum_{k: I_k\in\III_2} \Big[\big(\frac12-c'''\s^{1/4}
\big)\widetilde F^{(01)}_k+\s^\a \widetilde 
F^{(2)}_k\Big]\;.\nn\eea
Now, for $\s$ small, each term in square brackets is positive, simply 
because $\widetilde F^{(01)}_k\ge c\e$ and $\widetilde F^{(2)}_k\ge -4\e$,
so that (\ref{4.14a}) with the sums restricted to intervals of type 2 follows. 

\subsubsection{Intervals of type 3}

In this case we just use the fact that $||u_0-w||^2_{L^2(I_k)}\le
\int_{I_k}dy (2y)^2=(4/3)H_k^3$, simply because $u_0=w$ on the boundary
of $I_k$ and $|(u_0-w)'|\le 2$. Therefore, if $\s=\b\e^{-1/3}$ 
\be \bar c\b H_k^{1/2}||u_0-w||_{L^2(I_k)}\le c\b H_k^{2}\le
c'\b M^{-2}\le c''\s^{3/2} \widetilde F_k^{(01)}\;,\label{4.25}
\ee
where in the last inequality we used that 
$c (\b/\e)^{1/2}\le M\le c' 
(\b/\e)^{1/2}$, by Eq.(\ref{2.7b})-(\ref{4.3}), and $M^{-3}\le \h^{-1}
\widetilde F^{(01)}_k$, by the definition of type 3 interval. Using 
(\ref{4.25}), we get 
\bea  \sum_{k: I_k\in\III_3} \big(\widetilde F_k- \bar c\b 
H_k^{1/2}||u_0-w||_{L^2(I_k)}\big)&\ge& \frac12\sum_{k: I_k\in\III_3} 
\widetilde F_k^{(01)}+\big(1-\s^\a\big)\sum_{k: I_k\in\III_3} 
\widetilde F^{(2)}_k+\nn\\
&+&\sum_{k: I_k\in\III_3} \Big[\big(\frac12-c''\s^{3/2}
\big)\widetilde F^{(01)}_k+\s^\a \widetilde 
F^{(2)}_k\Big]\;.\nn\eea
Now, for $\s$ small, each term in square brackets is positive, simply 
because $\widetilde F^{(01)}_k\ge c\h\e\s^{-3/2}$
and $\widetilde F^{(2)}_k\ge -4\e$,
so that (\ref{4.14a}) with the sums restricted to intervals of type 3 follows. 

\subsubsection{Intervals of type 4}

In this case, if $H_{k^*}:=\max\{H_{k-1},H_k,H_{k+1}\}$, we have that 
$\max_{2k-2\le j\le 2k+4}\tilde h_{j}\ge H_{k^*}/3>2/M$. Therefore, 
\be \frac{\b c_0}7\sum_{j=2k-2}^{2k+4}\big(\tilde h_{j}-\frac1{M}
\big)^2\ge c\b H_{k^*}^2\ge \frac{c'\b}{M^2}\label{4.26}\ee
and, as a consequence, 
\be \widetilde F_4:=
\sum_{k: I_k\in\III_4}\widetilde F_k^{(01)}\ge c \b\sum_{k: I_k\in\III_4}
H_{k^*}^2\ge \frac{c'\b}{M^2}\NN_4
\quad \Rightarrow \quad \NN_4\le C\b^{-1}\widetilde F_4 M^2\;,
\label{4.27}\ee
with $\NN_4=|\III_4|$ the number of intervals of type 4. On the other hand, 
by Lemma 3, we have that 
\be \bar c\b \sum_{k: I_k\in\III_4} H_k^{1/2}||u_0-w||_{L^2(I_k)}
\le c\b \sum_{k: I_k\in\III_4} H_k^{3/2}||u_0-u_1||_{L^2(I_k)}^{1/3}\;.
\label{4.28}\ee
By Poincar\'e inequality, $||u_0-u_1||_{L^2(I_k)}^2\le \int_0^1dx \int_{I_k}
dy\, u_x^2\le 3\widetilde F_k^{(01)}$, so that 
\bea \bar c\b \sum_{k: I_k\in\III_4} H_k^{1/2}||u_0-w||_{L^2(I_k)}
&\le& c\b \sum_{k: I_k\in\III_4} H_k^{3/2}[\widetilde F_k^{(01)}]^{1/6}
\le\label{4.29}\\
&\le& c\b \Big[\sum_{k: I_k\in\III_4} H_k^{9/5}\Big]^{5/6}
\Big[\sum_{k: I_k\in\III_4} \widetilde F_k^{(01)}\Big]^{1/6}
\;,\nn\eea
where the last inequality is Minkowski's. Another application of Minkowski's 
inequality shows that 
\bea\Big[\sum_{k: I_k\in\III_4} H_k^{9/5}\Big]^{5/6}
&\le& \Big[\sum_{k: I_k\in\III_4} H_k^{2}\Big]^{3/4}
\Big[\sum_{k: I_k\in\III_4} 1\Big]^{1/12}\le \nn\\
&\le&
c\Big[\frac{\widetilde F_4}{\b}\Big]^{3/4}\NN_4^{1/12}
\le c'\Big[\frac{\widetilde F_4}{\b }\Big]^{3/4}
\Big[\frac{\widetilde F_4M^2}{\b }\Big]^{1/12}
\;,\label{4.30}
\eea
where in the last two inequalities we used (\ref{4.27}). Substituting in 
(\ref{4.29}) we find 
\be \bar c\b \sum_{k: I_k\in\III_4} H_k^{1/2}||u_0-w||_{L^2(I_k)}
\le c (\b M)^{1/6}\widetilde F_4 \le c'\s^{1/4}\widetilde F_4 
\;,\label{4.31}\ee
with $\s=\b\e^{-1/3}$. Eq.(\ref{4.31}) implies
\bea  \sum_{k: I_k\in\III_4} \big(\widetilde F_k- \bar c\b 
H_k^{1/2}||u_0-w||_{L^2(I_k)}\big)&\ge& \frac12\sum_{k: I_k\in\III_4} 
\widetilde F_k^{(01)}+\big(1-\s^\a\big)\sum_{k: I_k\in\III_4} 
\widetilde F^{(2)}_k+\nn\\
&+&\sum_{k: I_k\in\III_4} \Big[\big(\frac12-c'\s^{1/4}
\big)\widetilde F^{(01)}_k+\s^\a \widetilde 
F^{(2)}_k\Big]\;.\nn\eea
Now, for $\s$ small, each term in square brackets is positive, simply 
because $\widetilde F^{(01)}_k\ge c\e$
and $\widetilde F^{(2)}_k\ge -4\e$, so that (\ref{4.14a}) with the sums 
restricted to intervals of type 4 follows. \\

Combining the estimates for all different types of intervals, which are all 
valid for $\k$ and $\h\k^{-3}$ sufficiently small, we finally 
get (\ref{4.14a}), which implies Theorem 1, as discussed after (\ref{4.14a}).
\qed

\acknowledgments
We would like to thank S. Conti for an illuminating suggestion 
on the choice of the variational function $w$
and J. L. Lebowitz and E. H. Lieb for many useful discussions and for their 
encouragement on this project. Part of this work was carried out at the Max 
Planck Institute for Mathematics in Leipzig and at the Hausdorff Center for 
Mathematics in Bonn. A.G. acknowledges the Forschergruppe 
``Analysis and stochastics in complex physical systems'' and the 
ERC Starting Grant ``Collective Phenomena in Quantum and Classical
Many Body Systems" (CoMBoS-239694) for partial financial support.

\appendix\section{}\label{A}
\setcounter{equation}{0}
\renewcommand{\theequation}{\ref{A}.\arabic{equation}}

In this appendix we prove (\ref{2.4}). 
Without loss of generality, we can assume that $u_0(y)$ has a corner point in 
$y=0$. Now, for any fixed $\a\in(0,+\io)$ we rewrite:
\bea &&\int_0^h dy \int_{-\io}^{+\io} dy'\,
|u_0(y)-\tilde u_0(y')|^2 e^{-\a|y-y'|}=\frac4{\a}\int_0^h dy |u_0(y)|^2
-\nn\\
&&\hskip1.5truecm-2\lim_{N\to\io}\frac1{N}\int_{0}^{Nh}dy\int_{0}^{Nh}dy' 
\tilde u_0(y) \tilde u_0(y') e^{-\a|y-y'|}\;.\label{A.1}\eea
The latter integral is in a form suitable for applying the 
``Chessboard estimate with Dirichlet boundary conditions'' proved in 
\cite{GLL3}, see (3.12) of \cite{GLL3}. However, in this case we want
to use ``ferromagnetic'' reflections, rather than the ``antiferromagnetic'' 
reflections used in \cite{GLL3}: in other words, we want to keep 
reflecting $u_0$ around the locations of its corner points 
$y_i$, $i=0,1,\ldots,M_0-1$, {\it without changing sign to the reflected 
function}. The result, analogue to (3.12) in \cite{GLL3}, is:
\bea &&\frac1{N}\int_{0}^{Nh}dy\int_{0}^{Nh}dy' \tilde u_0(y) \tilde u_0(y')
e^{-\a|y-y'|}\le\nn\\
&&\hskip1.truecm \le \sum_{i=0}^{M_0-1}\int_{y_i}^{y_{i+1}}dy\int_{-\io}^{+
\io}dy' u_0^{(i)}(y) \tilde u_0^{(i)}(y')e^{-\a|y-y'|}\;,\label{A.2}\eea
with $u_0^{(i)}$ the restriction of $u_0$ to the 
interval $[y_i,y_{i+1}]$, and $\tilde u_0^{(i)}$ its periodic extension
to the whole real line. Eq.(\ref{A.2}), combined with (\ref{A.1}), gives 
(\ref{2.4}).\\

For completeness, we provide here a proof of (\ref{A.2}) along the lines of
\cite{GLL3} (and using a notation as close as possible to the one of 
\cite{GLL3}). We need to introduce some definitions.

{\bf Definition 1.} {\it Given a finite interval $[a,b]$ on the real line, 
let $\EE^\a_{a,b}:L^2([a,b])\to\RRR$ be the functional defined as
\be\EE^\a_{a,b}(w):=-\int_{a}^{b} dy \int_{a}^{b} dy'\, w(y)w(y')
e^{-\a|y-y'|}\;.\label{A.3}\ee}

{\bf Definition 2.} {\it Let $m,n\in\ZZZ^+\cup\{+\io\}$ 
be such that $m+n\ge 1$.
Let $\FF=\{f_{-m+1},\ldots,f_0,f_1,\ldots,f_n\}$
be a sequence of functions $f_i\in L^2([0,T_i])$ and $T_i>0$, 
with $-m< i\le n$. Let $z_{-m}=-\sum_{j=-m+1}^0 T_j$ 
and $z_i=z_{-m}+\sum_{j=-m+1}^i T_j$, for all $-m< i\le n$ (if $m=0$ it is 
understood that $z_0=0$). Then we define
$\f[\FF]\in L^2_{\rm loc}([z_{-m}, z_n])$ 
to be the function obtained by juxtaposing the functions $f_i$ on the real 
line, in such a way that, if $z_{i-1}\le y\le z_i$, then 
$\f[\FF](y)=f_i(y-z_{i-1})$, for all $i=-m+1,\ldots,n$.}

{\bf Definition 3.} 
{\it (i) Given $T>0$ and $f\in L^2([0,T])$, 
we define $\th f\in L^2([0,T])$ to be the reflection of $f$, 
namely $\th f(y)=f(T-y)$, for all $y\in[0,T]$.\\
(ii) If $f\in L^2([0,T])$, we define $\f[f]=\f[\FF_\infty(f)]\in L^2_{\rm loc}
(\RRR)$, where $\FF_\infty(f)=\{\ldots,f_0,f_1,\ldots\}$ is 
the infinite sequence with $f_n=\th^{n-1} f$.
\\
(iii) Given a sequence $\FF=\{f_{-m+1},\ldots,f_n\}$ as in Def.2,
we define $\FF_-=\{f_{-m+1},\ldots,f_0\}$ and $\FF_+=\{f_1,\ldots,f_n\}$
(if $m=0$ or $n=0$, it is understood that $\FF_-$ or, respectively, $\FF_+$ 
is empty) and we write $\FF=(\FF_-,\FF_+)$.\\
(iv) The reflections of $\FF_-$ and $\FF_+$ are
defined to be: $\th\FF_-=\{\th f_0,\ldots,\th f_{-m+1}\}$ and $\th\FF_+=
\{\th f_n,\ldots,\th f_1\}$.}\\

Given the definitions above, the analogue of the ``Chessboard estimate with 
Dirichlet boundary conditions'' of \cite{GLL3} adapted to the present context 
is the following.

{\bf Lemma A.1 [Chessboard estimate with Dirichlet boundary conditions]}
{\it Given a finite sequence of functions
$\FF=\{f_{1},\ldots,f_n\}$, $n\ge 1$, as in Definition 2, with $f_i\in L^2(
[0,T_1])$, we have:
\be \EE^{\a}_{0,z_n}\big(\f[\FF]\big)\ge \sum_{i=1}^n T_i e_\io(f_i)
\;,\label{A.4a}\ee
with 
\be e_\io(f_i):=\lim_{n\to\io}\frac{\EE^\a_{0,nT_i}(\f[f_i])}{nT_i}\;.
\label{A.5}\ee
(Note that the limit in the r.h.s. of (\ref{A.5}) exists, because $\f[f_i]$
is periodic and the potential $e^{-\a|y|}$ appearing in the definition 
of $\EE^\a_{0,nT_i}$ is summable).}\\

Lemma A.1 is the desired estimate. It immediately implies (\ref{A.2}). In fact,
let: (i) $n=NM_0$; (ii) $0=y_0<y_1<\cdots<
y_{n}=Nh$ be the locations of the corner points of 
$\tilde u_0$ in $[0,Nh]$; (iii) $T_i=y_i-y_{i-1}$; (iv)
$f_i(y)=u_0^{(i)}(y-y_{i-1})$, $i=1,\ldots,n$. With these definitions, 
$\f[\{f_1,\ldots,f_{n}\}]=\tilde u_0$ on $[0,Nh]$ and $e_\io(f_i)=
T_i^{-1}\int_{y_i}^{y_{i+1}}dy\int_{-\io}^{+
\io}dy' u_0^{(i)}(y) \tilde u_0^{(i)}(y')e^{-\a|y-y'|}$; in particular, 
(\ref{A.4a}) reduces to (\ref{A.2}). \\

We are then left with proving Lemma A.1. A basic ingredient in the proof of 
Lemma A.1 is the following ``reflection positivity estimate'' (which is the 
analogue of Lemma 1 of \cite{GLL3}).

{\bf Lemma A.2} {\it Given a finite sequence of functions 
$\FF=\{f_{-m+1},\ldots,f_0,f_1,\ldots,f_n\}=(\FF_-,\FF_+)$, as in 
Def.2 and 3, 
we have:
\be \EE^{\a}_{z_{-m},z_n}(\f[\FF])\ge \frac12\EE^{\a}_{-z_{n},z_n}
(\f[\FF_1])+\frac12\EE^{\a}_{z_{-m},-z_{-m}}(\f[\FF_2])\;,\label{A.8}\ee
where $\FF_1=(\th\FF_+,\FF_+)=\{\th f_n,\ldots,\th f_1,f_1,\ldots,f_n\}$ and 
$\FF_2=(\FF_-,\th\FF_-)=\{f_{-m+1},\ldots,f_0,\th f_0,\ldots,\th f_{-m+1}\}$.}
\\

{\cs Proof of Lemma A.2.} We rewrite
\bea  \EE^\a_{z_{-m},z_n}(\f[\FF])=&&-\int_{z_{-m}}^0 dy\int_{z_{-m}}^0 dy'
\f[\FF](y)\f[\FF](y')e^{-\a|y-y'|}\nn\\
&&-\int_{0}^{z_n} dy\int_{0}^{z_n} dy'
\f[\FF](y)\f[\FF](y')e^{-\a|y-y'|}\nn\\
&&-2\int_{z_{-m}}^0dy\int_0^{z_n}dy'
\f[\FF](y)\f[\FF](y')e^{-\a(y'-y)}\;.\label{A.6}\eea
Now, notice that last term on the r.h.s. of (\ref{A.6})
can be rewritten and estimated as:
\bea
&&\int_{z_{-m}}^0dy\int_0^{z_n}dy'
\f[\FF](y)\f[\FF](y')e^{-\a(y'-y)}\nn\\
&&=\int_{0}^{-z_{-m}}dy
\f[(\FF_-,\th\FF_-)](y)e^{-\a y}\int_0^{z_n}dy'
\f[(\th\FF_+,\FF_+)](y')e^{-\a y'}\label{A.7}\\
&& \le \frac12\Big[\int_{0}^{-z_{-m}}dy
\f[(\FF_-,\th\FF_-)](y)e^{-\a y}\Big]^2+\frac12\Big[
\int_0^{z_n}dy'\f[(\th\FF_+,\FF_+)](y')e^{-\a y'}\Big]^2\;,\nn\eea
which is equivalent to
\bea&& \int_{z_{-m}}^0dy\int_0^{z_n}dy'
\f[\FF](y)\f[\FF](y')e^{-\a(y'-y)}\le\nn\\ 
&&\qquad \le\frac12\int_{z_{-m}}^0dy\int_0^{-z_{-m}}dy'
\f[\FF_1](y)\f[\FF_1](y')e^{-\a(y'-y)}\label{A.7a}\\
&&\qquad\ +\frac12\int_{-z_{n}}^0dy\int_0^{z_{n}}dy'
\f[\FF_2](y)\f[\FF_2](y')e^{-\a(y'-y)}\;,\nn\eea
with $\FF_1=(\FF_-,\th\FF_-)$ and $\FF_2=(\th\FF_+,\FF_+)$.
Now, (\ref{A.8}) follows by plugging (\ref{A.7a}) into (\ref{A.6}) and by 
using that 
\bea &&-\int_{z_{-m}}^0 dy\int_{z_{-m}}^0 dy'
\f[\FF](y)\f[\FF](y')e^{-\a|y-y'|}\label{A.8a}\\&&
-\frac12\int_{z_{-m}}^0dy\int_0^{-z_{-m}}dy'
\f[\FF_1](y)\f[\FF_1](y')e^{-\a(y'-y)}=\frac12\EE^\a_{z_{-m},-z_{-m}}
(\f[\FF_1])\;,\nn\eea
and
\bea&&-\int_0^{z_n} dy\int_0^{z_{n}} dy'
\f[\FF](y)\f[\FF](y')e^{-\a|y-y'|}\label{A.9}\\&&
-\frac12\int_{-z_{n}}^0dy\int_0^{z_{n}}dy'
\f[\FF_2](y)\f[\FF_2](y')e^{-\a(y'-y)}=\frac12\EE^\a_{-z_{n},z_{n}}
(\f[\FF_2])\;.\nn\eea
\qed

At this point, in order to prove Lemma A.2, one needs to inductively iterate 
the key estimate (\ref{A.8}), as explained in the following.\\

{\cs Proof of Lemma A.1.}
We proceed by induction.\\
(i) If $n=1$, we first rewrite
\be \EE^{\a}_{0,2z_1}(\f[\{f_1,\th f_1\}])=
2\EE^{\a}_{0,z_1}(f_1) - 2\int_0^{z_1} dy 
\int_{z_1}^{2z_1} dy' f_1(y) \, \th f_1(y'-z_1) e^{-\a(y'-y)}
\;,\label{A.10}\ee
and we notice that, by definition of $\th f_1$,
the second term in the r.h.s. of (\ref{A.10}) can be rewritten and estimated as
\be \int_0^{z_1} dy 
\int_{z_1}^{2z_1} dy' f_1(y)f_1(2z_1-y') e^{-\a(y'-y)}=
\Big[\int_0^{z_1} dy f_1(y) e^{-\a(z_1-y)}\Big]^2\ge 0
\label{A.10a}\ee
By combining (\ref{A.10}) and (\ref{A.10a}) we get 
\be \EE^{\a}_{0,z_1}(f_1)\ge \frac12\EE^{\a}_{0,2z_1}(\f[\{f_1,\th f_1\}])\;.
\label{A.11}\ee
Iterating the same argument, we find:
\be \EE^{\a}_{0,z_1}(f_1)\ge \frac{\EE^{\a}_{0,2^mz_1}(\f[f_1^{\otimes 2^m}])}
{2^m}\;,\label{A.12}\ee
where, by definition, 
\be f_1^{\otimes 2^m}=\{\,\overbrace{f_1,\th f_1,\ldots,f_1,\th f_1}^{2^m\ 
{\rm times}}\,\}\;.\label{A.13}\ee
Taking the limit $m\to\io$ in (\ref{A.12}) we get the desired estimate: 
\be\EE^{\a}_{0,z_1}(f_1) \ge T_1 e_\io(f_1)\;.\label{A.13a}\ee
(ii) Let us now assume by induction that the bound is valid for all 
$1\le n\le k-1$,
$k\ge 2$, and let us prove it for $n=k$. There are two cases.\\

(a) $k=2p$ for some $p\ge 1$. If we reflect once, by Lemma A.2 we have:
\bea &&\EE^{\a}_{0,z_{2p}}(\f[\{f_1,\ldots,f_{2p}\}])\ge\nn\\
&&\quad\qquad\ge\frac12
\EE^{\a}_{0,2(z_{2p}-z_p)}(\f[\{\th f_{2p},\ldots,\th f_{p+2}, 
(\th f_{p+1})^{\otimes 2}, f_{p+2},\ldots f_{2p}\}])+\nn\\
&&\quad\qquad\ + \frac12 
\EE^{\a}_{0,2z_p}(\f[\{
f_1,\ldots,f_{p-1},f_p^{\otimes 2},\th f_{p-1},\ldots,\th f_1\}]) 
\label{A.14}\eea
If we now regard $(\th f_{p+1})^{\otimes 2}$ and $f_p^{\otimes 2}$
as two new functions in $L^2([0,2T_{p+1}])$ and in $L^2([0,2T_{p}])$,
respectively, the two terms in the r.h.s. of (\ref{A.14}) can be regarded as 
two terms with $n=2p-1$ and, by the induction assumption, they satisfy the 
bounds:
\bea && \EE^{\a}_{0,2(z_{2p}-z_p)}
(\f[\{\th f_{2p},\ldots,\th f_{p+2}, (\th f_{p+1})^{\otimes 2},f_{p+2},\ldots 
f_{2p}\}])\ge 2 \sum_{i=p+1}^{2p} T_i e_\io(f_i)\;,\nn \\
&& \EE^{\a}_{0,2z_p}(\f[\{f_1,\ldots,f_{p-1},f_p^{\otimes 2},\th f_{p-1},
\ldots,\th f_1\}])\ge 
2 \sum_{i=1}^{p} T_i e_\io(f_i)\;, \label{A.15}\eea
where we used that $e_\io((\th f_{p+1})^{\otimes 2})=e_\io(f_{p+1})$ and 
$e_\io(f_p^{\otimes 2})=e_\io(f_p)$. Therefore, the desired bound is proved.\\

(b) $k=2p+1$ for some $p\ge 1$. If we reflect once, by Lemma A.2 we have: 
\bea && \EE^{\a}_{0,z_{2p+1}}(\f[\{f_1,\ldots,f_{2p+1}\}])\ge\label{A.16} \\
&& \ge \frac12 
\EE^{\a}_{0,2(z_{2p+1}-z_{p+1})}(\f[\{\th f_{2p+1},\ldots,\th f_{p+3},
(\th f_{p+2})^{\otimes 2},f_{p+3},\ldots,f_{2p+1}\}])+\nn\\ 
&& \hspace{.1truecm}+\frac12 
\EE^{\a}_{0,2z_{p+1}}(\f[\{f_1,\ldots,f_p,f_{p+1}^{\otimes 2},\th f_p,\ldots,
\th f_1\}]) \nn\eea
The first term in the r.h.s. corresponds to $n=2p-1$ so by the induction 
hypothesis it is bounded below by $\sum_{i=p+2}^{2p+1} T_i 
e_\io(f_i)$. As regards the second term, using Lemma A.2 again, 
we can bound it from below by
\bea&& \frac14
\EE^{\a}_{0,2z_p}(\f[\{f_1,\ldots,f_{p},\th f_{p},\ldots,\th f_1\}])+\nn\\
&& +\frac14 
\EE^{\a}_{0,2z_{p}+4z_{p+1}}(\f[\{f_1,\ldots,f_{p},(f_{p+1})^{\otimes 4},
\th f_{p},\ldots,\th f_1\}])\label{A.17}\eea
By the induction hypothesis, the first term is bounded below by 
$(1/2)\sum_{i=1}^p T_i e_\io(f_i)$, and the second can be bounded
by Lemma A.2 again. Iterating we find:
\bea && \EE^{\a}(\f[\{f_1,\ldots,f_{2p+1}\}])\ge\label{A.18} \\
&& \quad \ge \sum_{i=p+2}^{2p+1} T_i e_\io(f_i)
+ \Big(\sum_{n\ge 1}2^{-n}\Big)\cdot \sum_{i=1}^p T_i e_\io(f_i)\Big)
+\nn\\
&&\hspace{2.truecm} +\lim_{n\to\io} 2^{-n} 
\EE^{\a}_{0,2z_{p}+2^m z_{p+1}}(\f[\{f_1,\ldots,f_{p},(f_{p+1})^{\otimes 2^m},
\th f_{p},\ldots,\th f_1\}])\;. \nn\eea
Note that the last term is equal to $T_{p+1} e_\io(f_{p+1})$, so (\ref{A.18})
is the desired bound. This concludes the proof of (\ref{A.9}). \qed


\end{document}